\documentclass[pra,superscriptaddress,aps,showpacs,showkeys,a4paper,nofootinbib,notitlepage]{revtex4-1}

\usepackage[utf8]{inputenc}
\usepackage[english]{babel}
\usepackage{amsmath}
\usepackage{amssymb}
\usepackage{amsfonts}
\usepackage{graphicx}
\usepackage{epstopdf}

\usepackage[T1]{fontenc}

\usepackage[tiny,center,uppercase]{titlesec}
\usepackage{xcolor}
\definecolor{link_blue}{RGB}{52,46,157}
\usepackage[colorlinks,citecolor=link_blue,linkcolor=link_blue,urlcolor=link_blue]{hyperref}
\usepackage[a4paper,textwidth=380pt,textheight=592pt,top=107pt,left=80pt]{geometry}

\newcommand{\ra}{\rangle}
\newcommand{\la}{\langle}

\DeclareMathOperator{\Tr}{Tr}
\DeclareMathOperator{\re}{Re}
\allowdisplaybreaks[2]

\begin{document}

\title{Physical background for parameters of the quantum Rabi model}

\author{I.\ D.\ Feranchuk}
\affiliation{Atomic Molecular and Optical Physics Research Group, Ton Duc Thang University, 19 Nguyen Huu Tho Str., Tan Phong Ward, District 7, Ho Chi Minh City, Vietnam}
\affiliation{Faculty of Applied Sciences, Ton Duc Thang University, 19 Nguyen Huu Tho Str., Tan Phong Ward, District 7, Ho Chi Minh City, Vietnam}
\affiliation{Belarusian State University, 4 Nezavisimosty Ave., 220030, Minsk, Belarus}
\author{A.\ V.\ Leonov}
\affiliation{Belarusian State University, 4 Nezavisimosty Ave., 220030, Minsk, Belarus}
\author{O.\ D.\ Skoromnik}
\email[Corresponding author: ]{olegskor@gmail.com}
\affiliation{Max Planck Institute for Nuclear Physics, Saupfercheckweg 1, 69117 Heidelberg, Germany}

\begin{abstract}
  We investigate the applicability of the two major approximations which are most commonly employed in the study of the quantum Rabi model, namely the description of a resonant cavity mode as a single-mode quantized field and the use of the rotating wave approximation. Starting from the Hamiltonian of a two-level system interacting with a multi-mode quantized field, we perform the canonical transformation of the field operators. This allows one to partition the Hamiltonian of the system into two parts. The first part is the interaction of the two-level system with a single collective field mode, while the second one describes the interaction with field fluctuations. The first part is usually associated with the resonant cavity mode. This division enables us to determine the applicability condition of the single-mode approximation. In addition we identify simple approximate relations for the description of the eigenstates, eigenfunctions and the time evolution of the quantum Rabi model beyond the rotating wave approximation.
\end{abstract}

\maketitle

\section{Introduction}\label{sec:intro}

The quantum Rabi model, describing the interaction between a two-level system and a single-mode quantized field \cite{ScullyB1997}, plays an extremely important role in various areas of physics, ranging from quantum optics \cite{KnightA1993,PhysRevA.88.043828,MandelB1995,KiffnerA2010} to nanomechanics \cite{PhysRevLett.96.127006} and trapped ions \cite{PhysRevA.92.033817}. Due to its simplicity and predictive ability this model has been extensively studied. Nevertheless, it still attracts a lot of theoretical interest, for example its integrability properties \cite{1751-8121-47-26-265303,JMathPhys.102104.2014,1751-8121-46-17-175301,1751-8121-46-41-415302,1751-8121-47-4-045301,JMathPhys.102104.2013,PhysRevLett.107.100401, PhysRevA.85.043805, PhysRevA.86.023822} or extension to the nonlinear regime \cite{PhysRevA.93.043814} were discussed only recently.

At the same time, quite often a further simplification of the quantum Rabi model is employed, when the quickly oscillating counter-rotating terms are neglected in the Hamiltonian - the Jaynes-Cummings model \cite{JaynesCummingsA1963}. Despite its remarkable success in the description of the atom-field interactions \cite{KnightA1993} a few extensions have been proposed, for example inclusion of the losses of the resonant mode through a lossy cavity \cite{PhysRevA.29.2627,PhysRevA.33.2444,PhysRevA.33.3610,PhysRevA.34.3077,PhysRevA.35.3433,PhysRevA.37.3175,PhysRevA.42.6884,PhysRevA.43.346,PhysRevA.44.4541,PhysRevA.44.6092,PhysRevA.47.2221} or the generalization to the case of a few discrete modes interacting with the atom \cite{PhysRevA.91.043840,PhysRevA.88.043816,PhysRevA.83.063806,PhysRevA.40.5116,PhysRevA.42.4336,PhysRevA.42.6873,PhysRevA.44.6043,PhysRevA.45.6610}. However, the influence of the multi-mode character of the quantized field on the atom dynamics was studied only lately \cite{PhysRevX.5.021035,PhysRevA.90.062104,PhysRevA.87.043817,PhysRevLett.110.100405,PhysRevLett.110.160401,0953-4075-46-22-224008,PhysRevA.83.053826,PhysRevA.77.043822,PhysRevA.69.043807}.

In the present paper we aim to analyze the two major approximations viz. the description of the quantized field as a single-mode and the use of the rotating wave approximation and to formulate simple analytical expressions which define the validity of the former approximation while allow the description of the system beyond the later one.

The single-mode approximation is related to the fact that in a real cavity the electromagnetic field is located in a limited volume within a finite time due to the energy dissipation processes. This expresses the situation that the quantum states of the electromagnetic field form a quasi-continuous spectrum and an atom in the cavity interacts with a wave packet of the field. The large number of field modes with different frequencies and directions of $\vec k$ vectors of this wave packet are centered around some resonant cavity mode. For this reason our goal is to establish the relation between the parameters of this wave packet and the parameters of the single-mode Rabi Hamiltonian. We want to briefly point here that a similar problem was recently solved for an electron, which moves in a field of a laser wave \cite{0953-4075-47-11-115601}.

As was mentioned above, the exact solution for the stationary states of the quantum Rabi model has been recently formulated \cite{PhysRevLett.107.100401}. However, it is expressed in a form of an infinite power series, hence the actual application of this solution in real calculations becomes problematic. Moreover, it describes the stationary states and not the time evolution of the system. For this reason, several alternative attempts have been performed to go beyond the rotating wave approximation, for example the development of a stochastic Schrödinger equation approach including cavity losses \cite{PhysRevA.90.023820}, the application of unitary transformations to the Rabi Hamiltonian \cite{0953-4075-46-3-035505}, the use of quasienergies and quasienergy states \cite{PhysRevLett.115.133601} or the construction of the symmetric and the generalized forms of the rotating wave approximation \cite{PhysRevA.84.042110,PhysRevA.87.033827}. For this reason, we investigated the Hamiltonian of the quantum Rabi model and were able to formulate extremely simple analytical expressions, which allow one to calculate the eigenvalues and eigenstates of the system with arbitrary accuracy.

Concluding, the paper is organized in the following way. In Sec.~\ref{sec:sec2} we employ the canonical transformation of the field operators, which allows us to sort out a single field mode and fluctuations with respect to the latter. The parameters of this collective field mode are defined through the parameters of the wave-packet. Moreover, the interaction of the two-level system with this collective field mode coincides with the quantum Rabi model. The interplay of the interaction of the atomic system with the collective field mode and the fluctuations determines the applicability of the single-mode approximation respectively. In Sec.~\ref{sec:sec3} we analyze the applicability of the rotating wave approximation within the framework of quantum Rabi model and demonstrate that the inclusion of the counter-rotating terms in the Rabi Hamiltonian changes predominantly the integrals of motion of the system. Next, we construct the approximate analytical expressions, which describe the stationary states and the time evolution of the system beyond the rotating wave approximation and demonstrate their validity in the whole range of variation of the coupling constant.

\section{Two level system in a multi-mode quantized field and transformations to the quantum Rabi model}\label{sec:sec2}

The system consisting of a two-level system, which interacts with a quantized electromagnetic field is described via the Schr\"{o}dinger equation
\begin{align}\label{eq:1}
  \hat H | \psi \rangle &= E | \psi \rangle,
\end{align}
where the Hamiltonian $\hat H$ consists of the three parts
\begin{align}
  \hat H &= \hat H_{\mathrm{a}} + \hat H_{\mathrm{f}} + \hat V_{\mathrm{af}}, \label{eq:2}
  \\
  \hat H_{\mathrm{a}} &= -  \frac{\epsilon}{2} | \chi_{\downarrow} \rangle \langle \chi_{\downarrow} | + \frac{\epsilon}{2} | \chi_{\uparrow} \rangle \langle \chi_{\uparrow}  |, \nonumber
  \\
  \hat H_{\mathrm{f}} &= \sum_{\vec{k},s} \omega_{\vec{k}} \hat{a}^{\dag}_{\vec{k},s} \hat{a}_{\vec{k},s}, \nonumber
  \\
  \hat V_{\mathrm{af}} &=- \frac{e_0}{m_e} \sum_{\vec{k},s}\left(\frac{2\pi}{V \omega_{\vec{k}}}\right)^{1/2} e^{i\vec{k}\cdot\vec{r}}(\vec{e}_{\vec{k},s}\cdot \hat{\vec{p}}) (\hat{a}^{\dag}_{-\vec{k},s} + \hat a_{\vec{k},s}). \nonumber
\end{align}
Here $\hat H_{\mathrm{a}}$ describes the atomic sub-system, $\hat H_{\mathrm{f}}$ the field sub-system and $\hat{V}_{\mathrm{af}}$ the interaction between the former and the latter, $\{|\chi_{\downarrow}\rangle, -\epsilon/2\}$ and $\{|\chi_{\uparrow}\rangle, \epsilon/2\}$ are the state vectors and the energies of the ground and exited states of the two-level system respectively, $e_0 < 0$ and $m_{\mathrm{e}}$ the charge and the mass of the electron, $\hat a^{\dag}_{\vec{k},s}, \hat a_{\vec{k},s}$ the creation and annihilation operators of the photon with the wave vector $\vec{k}$, the frequency $\omega_{\vec{k}} = |\vec{k}|$ and the polarization $\vec{e}_{\vec{k},s}$, $V$ the cavity volume and $\hat{\vec{p}}$ and $\hat{\vec{r}}$ the operators of the momentum and the coordinate acting in the Hilbert space of the atomic sub-system. We work in natural system of units in which $\hbar = c = 1$.

Let us discuss in more detail how the sums over the field states in Eq.~(\ref{eq:2}) are defined. These field states are centered around some frequency $\omega_{0}$ of an eigenmode of an ideal cavity. However, due to the energy dissipation processes in a real cavity this eigenmode becomes broadened and possesses width $\Delta\omega$ related to the relaxation time $\tau = 1 / \Delta\omega$ of the dissipation process, which in turn is associated with the cavity quality factor $Q = \omega_{0} / \Delta\omega = \omega \tau$. Consequently, the sums in Eq.~(\ref{eq:2}) consist of a macroscopic number of terms $N_{D}$, which we can relate to the parameters of the cavity
\begin{align}
  \label{eq:3}
  N_D = \frac{V}{(2\pi)^3} 4 \pi \omega_0^2\Delta \omega = \frac{V \omega_0^3}{\pi^2 Q},
\end{align}
where we consider that the cavity eigenmode is polarized.

Another important remark is to be done about the two-level description of the atomic system. It is evident that the representation of the atom as a two-level system instead of a many level system is valid only under the condition that the transition frequency $\epsilon$ between the states $|\chi_{\downarrow}\rangle$, $|\chi_{\uparrow}\rangle$ is close to the frequency $\omega_{0}$ and is highly detuned from the frequencies $\omega_{\lambda}$ of the other cavity eigenmodes and from the energy differences $\epsilon_{\mu}$ of the other atomic transitions:
\begin{eqnarray}\label{eq:4}
|\omega_0 - \epsilon| \ll |\omega_0 - \omega_{\lambda}|, \quad |\omega_0 - \epsilon| \ll |\epsilon - \epsilon_{\mu}|.
\end{eqnarray}
Consequently, in the following we will consider that the conditions defined via Eq.~(\ref{eq:4}) are fulfilled.

In order to derive the Hamiltonian $\hat{H}$ of the Rabi model from Eq.~(\ref{eq:2}) we will employ the dipole approximation $e^{i\vec{k}\cdot\vec{r}} \approx 1$ which has a good accuracy in the optical and radio frequency ranges. Then taking into account only two atomic transitions and the only one field polarization the Hamiltonian (\ref{eq:2}) can be transformed into the form
\begin{align}\label{eq:5}
  \hat H &= \frac{\epsilon}{2} \hat{\sigma}_3 + \sum_{\vec{k}} \omega_{\vec{k}} \hat{a}^{\dag}_{\vec{k} } \hat{a}_{\vec{k} } + \sum_{\vec{k}}\hat{\sigma}_1 M_{\vec{k}}(\hat{a}^\dag_{-\vec{k} } + \hat{a}_{\vec{k}}); 
  \\
  M_{\vec{k}} &= - \frac{e_0 }{m_{\mathrm{e}}}\left(\frac{2\pi}{V \omega_{\vec{k}}}\right)^{1/2}\langle \chi_{\uparrow}  |(\vec{e}_{\vec{k},s}\cdot\hat{\vec{p}})|\chi_{\downarrow}\rangle. \nonumber
\end{align}
Here Pauli matrices $\hat{\sigma}_{1,3}$ are acting in the atomic subspace $|\chi_{\downarrow}\rangle, |\chi_{\uparrow}\rangle$.

The Hamiltonian (\ref{eq:5}) still includes many field modes. In order to sort out the single-mode approximation we will employ the method of model Hamiltonians, which was introduced in the work \cite{BogolyubovB1989}. The main idea can be quickly formulated as follows: some model Hamiltonian is sorted out in the initial Hamiltonian in a way that it depends on a set of variational parameters, is exactly solvable and is a good approximation for the system under investigation. The variational parameters are to be adjusted such that this model Hamiltonian becomes the best possible approximation. This method was recently used in the problem of the interaction of a relativistic electron and a strong external laser field \cite{0953-4075-47-11-115601}. Consequently, in the present problem we introduce the model Hamiltonian for the Eq.~(\ref{eq:5}) as follows
\begin{align}\label{eq:6}
  \hat{H}_{\mathrm{A}} &= \frac{\epsilon}{2} \hat{\sigma}_3 + \sum_{\vec{k} < \Delta}\left[\tilde{\omega}_0 \hat{a}^\dag_{\vec{k}} \hat{a}_{\vec{k}} + \tilde{M}_0 \hat{\sigma}_1(\hat{a}_{\vec{k}} + \hat{a}^\dag_{-\vec{k}})\right] + \sum_{\vec{k} > \Delta} \omega_{\vec{k}} \hat{a}^\dag_{\vec{k}} \hat{a}_{\vec{k}},
  \\
  \hat{H} &\equiv \hat{H}_{\mathrm{A}} + \hat{H}_1 + \hat{H}_2. \nonumber
\end{align}
where the constant values $\tilde{\omega}_0$, $\tilde{M}_0$ and a small volume $\Delta$ in $\vec{k}$-space, centered around the vector $\vec k_{0}$ of the resonant mode are variational parameters of the model Hamiltonian and will be determined later \cite{BogolyubovB1989}. The sums $\sum_{\vec k < \Delta}$ and $\sum_{\vec k > \Delta}$ represent the summation inside and outside $\Delta$ region correspondingly and the operators $\hat H_{1,2}$ are found directly from Eq.~(\ref{eq:5})
\begin{align}\label{eq:7}
  \begin{aligned}
      \hat H_1 &=  \sum_{\vec k < \Delta}\left[ (\omega_{\vec{k}} - \tilde{\omega}_0) \hat{a}^\dag_{\vec{k}} \hat{a}_{\vec{k}} +  (M_{\vec{k}} - \tilde{M}_0)\hat{\sigma}_1 (\hat{a}_{\vec{k}}+\hat{a}^\dag_{-\vec{k}})\right], 
      \\
      \hat{H}_2 &= \sum_{\vec{k} > \Delta} M_{\vec{k}} \hat{\sigma}_1 (\hat{a}_{\vec k}+\hat{a}^\dag_{-\vec k}).
  \end{aligned}
\end{align}

By the definition the model Hamiltonian $\hat{H}_{\mathrm{A}}$ should quantitively describe the system, be diagonilizable and the perturbations due to the Hamiltonians $\hat{H}_{1,2}$ need to be small. For the diagonalization of $\hat{H}_{\mathrm{A}}$ let us utilize the method of canonical transformation, which was introduced by Bogolubov and Tyablikov in a polaron theory in the limit of a strong field \cite{BogolyubovA1950about,TyablikovA1951adyabatic,BogolyubovB2008}. For this purpose we go back to the coordinate representation of the field operators in Eq.~(\ref{eq:6})
\begin{align}\label{eq:8}
  \hat{H}_A &= \frac{\epsilon}{2} \hat{\sigma}_3 + \frac{1}{2}\tilde\omega_0 \sum_{\vec k<\Delta }    (\hat{p}_{\vec k}^2 + \hat{q}_{\vec k}^2) + \tilde M_0 \hat{\sigma}_1 \sqrt{2}\sum_{\vec k<\Delta }   \hat{q}_k + \sum_{\vec k > \Delta}\omega_{\vec k} \hat{a}^\dag_{\vec k} \hat{a}_{\vec k},
  \\
  \hat{q}_{\vec k} &= \frac{\hat{a}_{\vec k}+\hat{a}^\dag_{-\vec k}}{\sqrt{2}}, \quad \hat{p}_{\vec k} = - i \frac{\partial}{\partial q_{\vec k} } = - i \frac{\hat{a}_{\vec k} - \hat{a}^\dag_{-\vec k}}{\sqrt{2}}. \nonumber
\end{align}

Following Bogolubov \cite{BogolyubovB1989}, let us introduce the collective variable $\hat{D}$ in which all field modes are added coherently and the ``relative'' field variables $\hat{y}_{\vec k}$, which define quantum fluctuations relative to the collective mode
\begin{eqnarray}\label{eq:9}
  \hat{D} = \sum_{\vec k<\Delta }   \hat{q}_{\vec k}, \quad \hat{y}_{\vec k} = \hat{q}_{\vec k} - \frac{1}{N} \hat{D}, \quad \hat{q}_{\vec k} = \hat{y}_{\vec k} + \frac{1}{N} \hat{D}, \quad \sum_{k<\Delta }\hat{y}_{\vec k} = 0, \quad N = \sum_{\vec k<\Delta}1,
\end{eqnarray}
where $N$  is the number of modes in the volume $\Delta$ and in principle this number does not coincide with $N_D$ from Eq.~(\ref{eq:3}), however, being of the same order of magnitude.

The field momenta are calculated according to their definition \cite{LandauQM}:
\begin{eqnarray}\label{eq:10}
  \hat{p}_{\vec k} = - i \frac{\partial}{\partial q_{\vec k} } = - i \left\{\frac{\partial D }{\partial q_{\vec k}}\frac{\partial}{\partial D} + \sum_{\Delta \vec f}\frac{\partial y_{\vec f} }{\partial q_{\vec k} }\frac{\partial}{\partial y_{\vec f}}\right\}.
\end{eqnarray}

Evaluation of the derivatives with the help of Eq.~(\ref{eq:9}) gives the generalized momenta:
\begin{eqnarray}\label{eq:11}
  \hat{p}_{\vec k}  = \hat{\mathfrak{P}} + \hat{p}_{y_{\vec k}}, \quad  \sum_{\Delta k } p_{y_{\vec k}} = 0, \quad \hat{\mathfrak{P}} = - i \frac{\partial}{\partial D}, \quad \hat{p}_{y_{\vec k}} = - i \frac{\partial}{\partial y_{\vec k} } + \frac{i}{N} \sum_{\Delta \vec f}\frac{\partial}{\partial y_{\vec f}}.
\end{eqnarray}

Insertion of Eqs.~(\ref{eq:11}) and (\ref{eq:9}) into Eq.~(\ref{eq:8}) for the Hamiltonian leads to the separation of the collective coordinate $\hat D$, the fluctuation operators $\hat{y}_{\vec k}$ and the ``external'' variables $\hat{a}_{\vec k}$ and $\hat{a}^\dag_{\vec k}$, in which $\vec k > \Delta$:
\begin{align}\label{eq:12}
  \hat{H}_{\mathrm{A}} &= \frac{\epsilon}{2} \hat{\sigma}_3 + \frac{1}{2}\tilde \omega_0 \left[\frac{1}{N}\hat{D}^2 + N\hat{\mathfrak{P}}^2\right] +\tilde M_0 \hat{\sigma}_1 \sqrt{2} \hat{D} \nonumber
  \\
  &+ \frac{1}{2}\tilde\omega_0 \sum_{\vec k < \Delta  }(\hat{p}_{y_{\vec k}}^2 + \hat{y}_{\vec k}^2) + \sum_{\vec k > \Delta}\omega_{\vec k} \hat{a}^\dag_{\vec k} \hat{a}_{\vec k}.
\end{align}

Let us now quantize the collective and ``relative'' variables by introducing a new set of creation and annihilation operators
\begin{equation}\label{eq:13}
  \begin{aligned}
    \hat{D} &= \frac{\sqrt{N}}{\sqrt{2}} (\hat{A} + \hat{A}^\dag), \quad \hat{\mathfrak{P}} = -i \frac{1}{\sqrt{2 N}} (\hat{A} - \hat{A}^\dag), \quad [\hat{A}, \hat{A}^\dag] = 1,
    \\
    \hat{y}_{\vec k}  &= \frac{1}{\sqrt{2}}(\hat{\tilde b}_{\vec k} + \hat{\tilde b}_{\vec k}^\dag), \quad \hat{p}_{y_{\vec k}}  = - i \frac{1}{\sqrt{2}}( \hat{\tilde b}_{\vec k} -\hat{\tilde b}_{\vec k}^\dag),
    \\
   \hat{\tilde b}_{\vec k} &= \hat{a}_{\vec k} - \frac{1}{N}\sum_{\vec f < \Delta  } \hat{a}_{\vec f}, \quad [\hat{a}_{\vec k}, \hat{a}_{\vec k_{1}}^\dag] = \delta_{\vec k\vec k_1}, \quad [\hat{\tilde b}_{\vec k} ,\hat{\tilde b}_{\vec k_{1}}^\dag] = \delta_{\vec k\vec k_1} + \frac{1}{N}.
  \end{aligned}
\end{equation}

Then with the accuracy $1/N$ the Hamiltonian $\hat H_{\mathrm{A}}$, defined through Eq.~(\ref{eq:12}), transforms into the form
\begin{align}\label{eq:14}
  \hat{H}_{\mathrm{A}} &= \frac{\epsilon}{2} \hat{\sigma}_3 + \tilde\omega_0 \hat{A}^\dag \hat{A} + \tilde M_0 \hat{\sigma}_1 \sqrt{N}(\hat{A} + \hat{A}^\dag) \nonumber
  \\
  &+ \tilde\omega_0 \sum_{\vec k < \Delta  } \hat{\tilde{b}}_{\vec k}^\dag \hat{\tilde{b}}_{\vec k} + \sum_{\vec k > \Delta} \omega_{\vec k} \hat{a}^\dag_{\vec k} \hat{a}_{\vec k} \equiv \hat{H}_{\mathrm{QRM}} + \hat{H}_{\mathrm{f}} + \hat{H}_{\mathrm{e}},
  \\
  \hat{H}_{\mathrm{QRM}} &= \frac{\epsilon}{2} \hat{\sigma}_3  + \tilde\omega_0 \hat{A}^\dag \hat{A} + \tilde M_0 \hat{\sigma}_1 \sqrt{N}(\hat{A} + \hat{A}^\dag), \nonumber
  \\
  \hat{H}_{\mathrm{f}} &= \tilde\omega_0 \sum_{\vec k < \Delta  }   \hat{\tilde{b}}_{\vec k}^\dag \hat{\tilde{b}}_{\vec k}, \nonumber
  \\
  \hat{H}_{\mathrm{e}} &= \sum_{\vec k > \Delta}\omega_{\vec k} \hat{a}^\dag_{\vec k} \hat{a}_{\vec k}, \nonumber
\end{align}
where the normal ordering for operators is assumed and the energy of ``vacuum oscillations'' is not taken into account. In this representation the operator $\hat{H}_{\mathrm{QRM}}$ which corresponds to the single-mode approximation viz. quantum Rabi model is completely separated from the operators $\hat{H}_{\mathrm{f}}$ and $\hat{H}_{\mathrm{e}}$ describing fluctuations relative to the resonant cavity mode with the frequency $\omega_{0}$ and the external field modes are not included in the wave packet respectively.

Consequently, one of the conditions to be satisfied for the model Hamiltonian is fulfilled, namely the operator $\hat H_{\mathrm{A}}$ can be exactly diagonalized, and therefore, the state vector of the system is represented as the product:
\begin{align}\label{eq:15}
|\Psi \ra = |\Psi_{\mathrm{QRM}}\ra |\{n_{\mathrm{f}}\}\ra |\{n_{\mathrm{e}}\}\ra ,  \quad  \hat{\tilde{b}}_{\vec k}^\dag \hat{\tilde{b}}_{\vec k} |n_{\vec k}^{\mathrm{f}}\ra  =  n_{\vec k}^{\mathrm{f}} |n_{\vec k}^{\mathrm{f}}\ra, \quad \hat{a}_{\vec k}^\dag \hat{a}_{\vec k} |n_{\vec k}^{\mathrm{e}}\ra  = n_{\vec k}^{\mathrm{e}}|n_{\vec k}^{\mathrm{e}}\ra,
\end{align}
where the state vector $|n_{\vec k}^{\mathrm{f}}\ra$ defines the state of the ``fluctuations'' and the state vector $|n_{\vec k}^{\mathrm{e}}\ra$ the state of the non-resonant electromagnetic field not interacting with an atom.  The remaining contribution $|\Psi_{\mathrm{QRM}}\ra$ describes the state of the atom interacting with a collective resonant mode of the field and satisfies the equation, which is up to the notations $\tilde\omega_{0}$ and $\tilde M_{0}$ completely equivalent to the conventional equation \cite{ScullyB1997} for the single mode quantum Rabi model:
\begin{eqnarray}\label{eq:16}
  \hat{H}_{\mathrm{QRM}}|\Psi_{\mathrm{QRM}}\ra = \left[\frac{\epsilon}{2} \hat{\sigma}_3 + \tilde\omega_0 \hat{A}^\dag \hat{A} +\tilde M_0 \hat{\sigma}_1 \sqrt{N}(\hat{A} + \hat{A}^\dag)\right]|\Psi_{\mathrm{QRM}}\ra = E |\Psi_{\mathrm{QRM}}\ra.
\end{eqnarray}

In what follows we will estimate the different terms in the Hamiltonian in Eq.~(\ref{eq:16}). However, let us note that in order the Eq~(\ref{eq:5}) to be fulfilled the difference $\omega_{0} - \tilde{\omega}_{0}$ needs to be small and consequently
\begin{align}
  \label{eq:17}
  \frac{\omega_{0} - \tilde{\omega}_{0}}{\omega_{0}} \approx \frac{\Delta \omega}{\omega_{0}} = O\left(\frac{1}{Q}\right).
\end{align}
For this reason below we will drop tilde on top of all quantities, until the very end of this section.

Quite often the Hamiltonian operator $\hat{H}_{\mathrm{QRM}}$ is written in the dimensionless form, when a special system of units is employed, i.e. $\omega_{0} \rightarrow 1$ and $\epsilon \rightarrow \omega_0 \bar{\epsilon}; \ E \rightarrow \omega_0 \bar{E}$. In this dimensionless form Eq.~(\ref{eq:16}) reads as
\begin{eqnarray}\label{eq:18}
  \hat{H}_{\mathrm{QRM}} = \frac{\bar{\epsilon}}{2} \hat{\sigma}_3 + \hat{A}^\dag \hat{A} + f \hat{\sigma}_1 (\hat{A} + \hat{A}^\dag).
\end{eqnarray}

As mentioned already above the Hamiltonian $\hat{H}_{\mathrm{QRM}}$ defined by Eq.~(\ref{eq:16}) coincides with the conventional form of the Hamiltonian of the quantum Rabi model. However, the operators $\hat{A}$ and $\hat{A}^\dag$ describe the collective field mode and not the resonant cavity mode as in the usual case. Moreover, the atom is coupled to this collective field mode through the constant $f$ which is also different and depends on both the atom and on the cavity parameters. For this reason, let us determine the contribution of the additional terms, contained in the total Hamiltonian operator (\ref{eq:14}), assuming that the initial state of the electromagnetic field is the wave packet
\begin{align}\label{eq:21}
  |\Psi_{\mathrm{f}} \ra = \exp \left\{ \sum_{\vec k} \left[ u_{\vec k}\hat{a}^\dag_{\vec k}  - u^*_{\vec k}\hat{a}_{\vec k}\right]\right\} | 0 \ra, \quad  \hat{a}_{\vec k}  | 0 \ra = 0,
\end{align}
which is a set of coherent states with the parameters $u_{\vec k}$. This wave packet is centered around the resonant cavity mode, with the frequency $\omega_{0}$, in the $k$-space and will be modeled with the Gaussian distribution
\begin{align}\label{eq:22}
  u_{\vec k} &= C\exp\left\{- \frac{\vec k^2_{\bot}}{2\varkappa_1^2\omega_0^2}\right\} \exp\left\{- \frac{(\omega - \omega_0)^2}{2\varkappa_2^2\omega_0^2}\right\},
  \\
  \vec k &= \vec k_{\bot} + \omega \frac{\vec k_0}{\omega_0}, \quad \vec k_{\bot}\cdot\vec k_0 = 0. \nonumber
\end{align}
The two dimensionless parameters $\varkappa_1 = (\omega_0 S)^{-1}$ and $\varkappa_2 = \Delta \omega/\omega_0 = Q^{-1} $ are the angular and the frequency spreads correspondingly and $S$ is the cavity transverse area.

The constant $C$ in Eq.~(\ref{eq:22}) is deduced from the normalization of the state $|\Psi_{\mathrm{f}}\rangle$ on the total energy $W$ of the resonant mode in the cavity:
\begin{align}\label{eq:23}
  W &= \langle \Psi_{\mathrm{f}}|  \sum_{\text{all }\vec k} \omega_{\vec k} \hat{a}^\dag_{\vec k} \hat{a}_{\vec k} |\Psi_{\mathrm{f}} \rangle \approx \omega_{0}\langle \Psi_{\mathrm{f}}|  \sum_{\text{all }\vec k} \hat{a}^\dag_{\vec k} \hat{a}_{\vec k} |\Psi_{\mathrm{f}} \rangle \nonumber
  \\
    &= C^2 \frac{\omega_{0}V}{(2\pi)^3}\int d\omega d \vec k_{\bot} |u_{\vec k}|^2  = C^2 \frac{V}{ 8\pi^3 }\omega_0^4 \pi^{3/2}\varkappa_1^2\varkappa_2 \Rightarrow \nonumber
  \\
  C &= \sqrt{\frac{8\pi^{3/2}W}{V \varkappa_1^2 \varkappa_2 \omega_0^4}}.
\end{align}

As all relevant quantities of the field wave packet are defined, we can proceed with the estimation of the different terms in the Hamiltonian $\hat{H}_{\mathrm{A}}$ in Eq.~(\ref{eq:14}).

The first term defines the contribution to the energy of the collective field mode $\omega_{0}\hat{A}^{\dag}\hat{A}$:
\begin{align}\label{eq:24}
  E_{0} = \langle \Psi_{\mathrm{f}}| \omega_0  \hat{A}^\dag \hat{A}|\Psi_{\mathrm{f}} \rangle &\approx  \langle \Psi_{\mathrm{f}}|  \sum_{\vec k < \Delta} \omega_0 \hat{a}^\dag_{\vec k} \hat{a}_{\vec k} |\Psi_{\mathrm{f}} \rangle = \frac{\omega_0 C^2 V}{(2\pi)^3}\int d\vec k |u_{\vec k}|^2 \nonumber
  \\
                                                                                      &= \frac{8 \pi^{3/2} W 2^3}{(2\pi)^3}\left(\int_0^{\frac{\Delta_1}{\varkappa_1 \omega_0}}dz e^{-z^2}\right)^2 \int_0^{\frac{\Delta_2}{\varkappa_2 \omega_0}}dz e^{-z^2}\nonumber
  \\
                                                                                      &= W \Phi^3(\delta),
\end{align}
where we assumed that the volume $\Delta$ in $k$-space can be parametrized as $\Delta = \Delta_1^2 \Delta_2 = \delta^3 \varkappa_1^2\varkappa_2 \omega_0^3$. Here $\delta \sim 1$ is the dimensionless parameter depending on the particular resonator form and $\Phi(z) = 2/\sqrt{\pi}\int_0^z e^{- t^2} dt$ is the error function.

The remaining terms in $\hat{H}_{\mathrm{A}}$ are estimated in a similar way:
\begin{align}
  N &= \sum_{\vec k < \Delta}1 = \frac{V}{(2\pi)^3}\delta^3 \varkappa_1^2\varkappa_2 \omega_0^3, \label{eq:25}
  \\
  E_{\mathrm{f}} &= \langle \Psi_{\mathrm{f}}|\hat H_{\mathrm{f}}|\Psi_{\mathrm{f}} \rangle =  \langle \Psi_{\mathrm{f}}|  \sum_{\vec k < \Delta}  \left[\hat{a}^\dag_{\vec k}- \frac{1}{N}\sum_{\vec f < \Delta}\hat{a}^\dag_{\vec f}\right]  \left[\hat{a}_{\vec k}- \frac{1}{N}\sum_{\vec f < \Delta}\hat{a}_{\vec f}\right] |\Psi_{\mathrm{f}} \rangle \nonumber
  \\
    &=C^2\sum_{\vec k < \Delta}  \left[u^*_{\vec k}- \frac{1}{N}\sum_{\vec f < \Delta}u^*_{\vec f}\right]  \left[u_{\vec k}- \frac{1}{N}\sum_{\vec f < \Delta}u_{\vec f}\right] = \nonumber
  \\
    &= C^2 \frac{V}{(2\pi)^3}  \left(\int d \vec k |u_{\vec k}|^2 - \frac{V}{(2\pi)^3 N} \left| \int d \vec k  u_{\vec k} \right|^2\right) \nonumber
  \\
    &= W \Phi^3(\delta)\left(1-\frac{2^3 \pi^{\frac{3}{2}}}{\delta^3}\frac{\Phi^6(\frac{\delta}{\sqrt{2}})}{\Phi^3(\delta)}\right), \label{eq:26}
  \\
  E_{\mathrm{e}} &= \langle \Psi_{\mathrm{f}}| \hat H_{\mathrm{e}}|\Psi_{\mathrm{f}} \rangle \approx \langle \Psi_{\mathrm{f}}|  \sum_{\vec k >\Delta} \omega_0 \hat{a}^\dag_{\vec k} \hat{a}_{\vec k} |\Psi_{\mathrm{f}} \rangle = W \left(1 - \Phi^3(\delta)\right), \label{eq:27}
  \\
  E_{\mathrm{a-f}} &= \langle \Psi_{\mathrm{f}}| M_0 \hat{\sigma}_1 \sum_{\vec k < \Delta}(\hat{a}_{\vec k}^{\dag}+\hat{a}_{\vec k})|\Psi_{\mathrm{f}} \rangle \approx 2M_{0} C\frac{V}{(2\pi)^{3}}\int d\vec k u_{\vec k} \nonumber
  \\
    &= \frac{4\sqrt{2}C \pi^{3/2}}{\delta^{3}}N M_{0} \Phi^{3}\left(\frac{\delta}{\sqrt{2}}\right) = 8\sqrt{N}\sqrt{\frac{\pi^{\frac{5}{2}}}{\delta^{3}}\frac{W}{V}}\frac{e_{0}}{m_{\mathrm{e}}\omega_{0}}|\langle \chi_{\uparrow}  |(\vec{e}_{\vec k_{0},s}\cdot\hat{\vec{p}})|\chi_{\downarrow}\rangle|. \label{eq:28}
\end{align}

Now one can use the freedom in the choice of the variational parameters in the model Hamiltonian $\hat{H}_{\mathrm{A}}$. The first parameter $\delta$ can be chosen in a way that the average contribution $E_{\mathrm{f}} = \langle \Psi_{\mathrm{f}}|\hat{H}_{\mathrm{f}}|\Psi_{\mathrm{f}} \rangle $ of the fluctuations relative to the collective field mode is equal to zero. This corresponds to the determination of $\delta$ from the solution of the equation
\begin{align}
  \left(1-\frac{2^3 \pi^{\frac{3}{2}}}{\delta^3}\frac{\Phi^6(\frac{\delta}{\sqrt{2}})}{\Phi^3(\delta)}\right) = 0, \quad \delta = 3.54.\label{eq:29}
\end{align}
We also note, that the particular value of the parameter $\delta$ depends on the actual distribution of the wave packet modes, but in every case its value can be calculated in an analogous way.

The contribution of the external modes can be neglected, as the ratio of $E_{\mathrm{e}} / E_{0}$ is 
\begin{align}\label{eq:30}
  \frac{E_{\mathrm{e}}}{E_{0}} = \frac{\left(1 - \Phi^3(\delta)\right)}{\Phi^{3}(\delta)} \approx 1.64\times 10^{-6}.
\end{align}

Next, we pay attention to the fact that even if the average value of $\hat{H}_{\mathrm{f}}$ is equal to zero the variance of $\hat{H}_{\mathrm{f}}^2$ is not equal to zero and influences the validity of the single-mode approximation. In order to estimate this variance, we consider the field as a noninteracting photon gas. Consequently, we can use the well known estimation from statistical mechanics \cite{LandauStat}, i.e. the variance is proportional to the $\sqrt{N}$, therefore
\begin{align}\label{eq:31}
  \mathcal{D}E_{\mathrm{f}} = \sqrt{\langle\hat{H}_{\mathrm{f}}^{2}\rangle} = \omega_{0}\sqrt{N}.
\end{align}

As all estimations have been performed we can conclude that \textit{in order to determine the validity of the single-mode approximation} we, therefore, should compare the interaction energy between the atom and the field $E_{\mathrm{a-f}}$ with the variance of the fluctuations $\mathcal{D}E_{\mathrm{f}}$:
\begin{align}\label{eq:32}
  \mu &\equiv \frac{E_{\mathrm{a-f}}}{\mathcal{D}E_{\mathrm{f}}} = 8\sqrt{\frac{\pi^{\frac{5}{2}}}{\delta^{3}}}\sqrt{\frac{W}{V}}\frac{e_{0}}{m_{\mathrm{e}}\omega_{0}^{2}}|\langle \chi_{\uparrow}  |(\vec{e}_{\vec{k},s}\cdot \hat{\vec{p}})| \chi_{\downarrow} \rangle| \gtrsim 1.
\end{align}
In order to estimate the transition matrix element we express the velocity of the electron through the transition frequency and since $\bar \epsilon$ and $\omega_{0}$ are of the same order of magnitude, we can write according to the virial theorem that $\omega_{0} = m_{\mathrm{e}} v^{2}_{\mathrm{e}} / 2$ \cite{LandauQM}. Consequently, the final equation for the parameter $\mu$ reads
\begin{align}\label{eq:33}
  \mu &\approx 7 e_{0}\sqrt{\frac{W}{V}\frac{1}{m_{\mathrm{e}}\omega_{0}^{3}}} \gtrsim 1.
\end{align}

From the previous equation, we can immediately conclude that the applicability condition of the single-mode approximation highly depends on the frequencies of the resonant modes of the cavity. The estimation of the critical energy density $w_{\mathrm{c}} = W_{\mathrm{c}}/V$:
\begin{align}\label{eq:34}
  w_{\mathrm{c}} = \frac{m_{\mathrm{e}}\omega_{0}^{3}}{49 e_{0}^{2}} \approx \frac{5.7\times 10^{10}}{\lambda_{0}^{3}[\mathrm{nm}]}\left[\frac{\mathrm{J}}{\mathrm{cm}^{3}}\right],
\end{align}
shows that for the optical frequencies range the energy density inside the cavity should be greater than $w_{\mathrm{c}} \gtrsim 10^{2}\, \mathrm{J}/\mathrm{cm}^{3}$, while for the radio frequencies range $w_{\mathrm{c}}\gtrsim 1\, \mathrm{nJ} / \mathrm{cm}^{3}$.

We have investigated the two out of three conditions to be satisfied by the model Hamiltonian $\hat{H}_{\mathrm{A}}$, namely its diagonalization and physical meaning. However, during the estimations above, the variational parameters $\tilde \omega_{0}$ and $\tilde M_{0}$ were replaced by the corresponding parameters of the resonant mode of the cavity. For this reason, in order to conclude our formulation, the connection between $\omega_{0}$, $M_{0}$ and $\tilde\omega_{0}$, $\tilde M_{0}$ should be established. Consequently, we construct the first-order perturbation theory which will determine the corrections due to the $\hat{H}_{1}$ and $\hat{H}_{2}$:
\begin{eqnarray}\label{eq:35}
  \left\{\hat H_{\mathrm{A}} - E \right\}  |\Psi\ra = - (\hat H_1 + \hat H_2)|\Psi\ra,
\end{eqnarray}

The formal introduction of the small parameter $\lambda$ on the right hand side of Eq.~(\ref{eq:35}) and the corresponding expansion of the state vector $|\Psi\rangle$ and the eigenvalue $E$ in a power series in $\lambda$
\begin{align*}
  |\Psi\ra = |\Psi^0\ra + \lambda |\Psi^1\ra, \quad E = E^0 + \lambda E^1,
\end{align*}
yields
\begin{eqnarray}\label{eq:36}
  (E^0 + \lambda E^1 - \hat H_{\mathrm{A}})(|\Psi^0\ra + \lambda |\Psi^1\ra) = \lambda(\hat H_1 + \hat H_2)(|\Psi^0\ra + \lambda |\Psi^1\ra).
\end{eqnarray}
The equality of terms with the same powers of $\lambda$ determines the corresponding corrections to the state vector and the eigenvalue
\begin{equation}\label{eq:37}
  \begin{aligned}
    &\lambda^0:\quad (E^0 - \hat H_{\mathrm{A}})|\Psi^0\ra = 0,
    \\
    &\lambda^1:\quad E^1 |\Psi^0\ra + (E^0 - \hat H_{\mathrm{A}})|\Psi^1\ra = (\hat H_1 + \hat H_2)|\Psi^0\ra.
  \end{aligned}
\end{equation}

Consequently, we conclude that the first equation at the zeroth power of $\lambda$ leads to the Hamiltonian of quantum Rabi model Eq.~(\ref{eq:16}), while the second determines the first correction $E^{1}$ to the energy of the system
\begin{eqnarray}\label{eq:38}
  E^1 = \la\Psi^0| (H_1+H_2)|\Psi^0\ra.
\end{eqnarray}

As was demonstrated in the reference \cite{FeranchukB2015}, the optimal values for the two unknown quantities $\tilde \omega_0$ and $\tilde M_0$ in the model Hamiltonian $\hat H_{\mathrm{A}}$ can be found from the condition
\begin{align}\label{eq:39}
  E^{1} = 0.
\end{align}
However, the calculation of the expectation value $\la\Psi^0| (\hat H_1 + \hat H_2)|\Psi^0\ra$ is rather cumbersome and was described in great detail in our paper \cite{0953-4075-47-11-115601}. Therefore, we will not present this calculations here, however, we note that the condition (\ref{eq:39}) leads to the following result
\begin{eqnarray}\label{eq:40}
  \tilde \omega_0 = \frac{1}{N}\sum_{\vec k<\Delta}\omega_{\vec k}; \quad \tilde M_0 = \frac{1}{N}\sum_{\vec k < \Delta}M_{\vec k}. 
\end{eqnarray}

The Eqs.~(\ref{eq:40}) allow a simple physical interpretation, namely the optimal choice of the parameters of the single collective field mode corresponds to the average frequency of the wave packet, which in turn depends on the cavity Q-factor.

Moreover, in the work \cite{0953-4075-47-11-115601} the influence of the second-order corrections on the validity of the single-mode approximation was investigated and it was shown that under the condition defined by Eq.~(\ref{eq:32}) this contribution can be omitted.

\section{Analysis of the quantum Rabi model beyond the rotating wave approximation}
\label{sec:sec3}

In the previous section we demonstrated that under the conditions defined by Eqs.~(\ref{eq:4}) and (\ref{eq:33}) the problem of the interaction between a two-level system and a single-mode qunatized electromagnetic field in the cavity is reduced to the solution of the following Schr\"{o}dinger equation
\begin{align}\label{eq:41}
  \hat{H}_{\mathrm{QRM}} |\psi \rangle &= E |\psi \rangle,
  \\
  \hat{H}_{\mathrm{QRM}} &= \frac{\epsilon}{2} \hat{\sigma}_3 + \hat{A}^\dag \hat{A} + f \hat{\sigma}_1 (\hat{A} + \hat{A}^\dag) \nonumber
\end{align}
with the coupling constant $f$ defined as
\begin{align}\label{eq:42}
  f &= \frac{M_{0}}{\omega_{0}}\sqrt{N} = - \frac{e_0 }{m_{\mathrm{e}}} \langle \chi_{\uparrow}  |(\vec{e}_{\vec{k}_{0},s}\cdot \hat{\vec{p}})| \chi_{\downarrow} \rangle \sqrt{\frac{2\pi N }{V\omega_{0}^{3}}} \nonumber
  \\
  &\approx - e_{0}v_{\mathrm{e}}\sqrt{\frac{2\pi N_{\mathrm{D}} }{V\omega_{0}^{3}}} = -e_{0}v_{\mathrm{e}}\sqrt{\frac{2}{\pi Q}}.
\end{align}
and we also dropped bar on the top of $\epsilon$. Here we replaced for simplicity the Gaussian form of the wave packet with the rectangular one and introduced the cavity quality factor $Q$ and the electron velocity $v_{\mathrm{e}}$.
 
In modern applications the dimensionless coupling constant $f$ and the field amplitude $\sqrt{\bar n}$ ($\bar n = \langle\hat A^{\dag}\hat A\rangle$ being the average number of photons of the resonant mode) are varied in the very broad ranges \cite{HarocheB2006,0034-4885-69-5-R02,PhysRevLett.68.1132}. In particular, even for the relatively small $f$ but in the strong electromagnetic field, the ratio $\xi = \Omega/\omega_0 = 2f\sqrt{\bar n}$ of the Rabi frequency $\Omega$ to the field frequency $\omega_0$ can be quite large \cite{ChiorescuA2004}. For example, in recent experiments with superconducting qubits the strong coupling limit has been reached \cite{SciRep.6.26720.2016,PhysRevLett.115.133601} and in many papers it was demonstrated that the numerical analysis of the experimental data requires the solution beyond the rotating wave approximation \cite{Feranchuk2009517,PhysRevA.80.063826,Browne:2000-07-10T00:00:00:0950-0340:1307}. In addition the strong driving field can be also used for an effective control of a two-level system \cite{PhysRevLett.96.107001,Shim201487,Saiko2008,0953-4075-47-15-155502,PhysRevLett.112.010502}. Moreover, when $\bar n \gg 1$ the electromagnetic field is usually considered classically, and it is relevant to completely describe the influence of quantum effects and the validity of the rotating wave approximation in this limit \cite{ChiorescuA2004}. Furthermore, we note here that similar effects can arise when an electron moves in a strong electromagnetic wave, where the quantum fluctuations can lead to the collapse--revival dynamics \cite{PhysRevA.87.052107}.

As was demonstrated in the work \cite{Feranchuk2009517}, the validity of the rotating wave approximation for the solution of the Schr\"{o}dinger equation is defined through the inequality
\begin{eqnarray}\label{eq:43}
  \xi = \frac{\Omega}{\omega_0}=2f\sqrt{\bar n} \ll 1,
\end{eqnarray}
which, with the help of Eq.~(\ref{eq:42}), can be cast into the form
\begin{align}\label{eq:44}
  \xi = 2|e_{0}|v_{\mathrm{e}}\sqrt{\frac{2}{\pi}\frac{W}{Q \omega_{0}}} \ll 1.
\end{align}

Consequently, when the parameter $\xi$ is larger than unity the system description should be performed beyond the rotating wave approximation. There is only one exception, namely when the cavity eigenmode is circularly-polarized and the counter-rotating terms in Eq.~(\ref{eq:41}) are identically equal to zero \cite{ScullyB1997}.

Concluding, the development of the effective numerical and analytical methods for the description of the evolution of the quantum Rabi model that is valid within the whole range of the variation of the parameters is an actual problem. The approximation of the stationary states, which is uniformly convergent to the exact numerical results in the whole range of variation of coupling constants of this model was introduced in references \cite{0305-4470-29-14-026,PhysRevLett.99.173601,*PhysRevLett.99.259901}, while in the work \cite{Feranchuk2011385} an analogous approach for the description of the quantum evolution operator beyond the rotating wave approximation was presented. This allowed one to provide the theoretical description for the new effects that can appear in the regime $\xi \gg 1$, i.e. beyond the rotating wave approximation in the evolution of the quantum Rabi model, namely the suppression of the collapse--revival effect \cite{PhysRevA.93.063834,PhysRevLett.44.1323,PhysRevA.23.236,PhysRevLett.76.1800,Feranchuk2009517} and the qualitative changes of the time evolution of the population difference in comparison with the Rabi oscillations \cite{Feranchuk20094113}. These effects are justified by the experimental data introduced in the reference \cite{ChiorescuA2004}. Consequently, below we will consider briefly some of these results.

It is well-known that the exact solution of the evolution problem for any quantum system is defined by the eigenfunctions of the corresponding Hamiltonian. Let us split the Hamiltonian of the quantum Rabi model (\ref{eq:41}) into two parts:
\begin{align}
  \hat{H}_{\mathrm{QRM}} &\equiv \hat{H}_0 + \hat{H}_1,\label{eq:45}
  \\
  \hat{H}_0 &= \frac{\epsilon}{2} \hat{\sigma}_3 + \hat{A}^\dag\hat{A}+f\left( \hat{\sigma}_+ \hat{A} + \hat{\sigma}_- \hat{A}^\dag \right), \label{eq:46}
  \\
  \hat{H}_1 &= f\left( \hat{\sigma}_+ \hat{A}^\dag + \hat{\sigma}_- \hat{A}\right), \label{eq:47}
\end{align}
where the operator $\hat{H}_0$ corresponds to the rotating wave approximation.

Here we stress, that the quantum Rabi model possesses the following exact integral of motion, which is called the combined parity \cite{0305-4470-29-14-026}
\begin{eqnarray}\label{eq:48}
  \hat{P} = \hat{\sigma}_3 \hat{S}, \quad \hat{S}=e^{i \pi \hat{A}^\dag \hat{A} },
\end{eqnarray}
and is substantially different from the one $\hat{J}=\left( \frac{1}{2}\hat{\sigma}_3 +\hat{A}^\dag \hat{A} \right)$ corresponding to the rotating wave approximation. Consequently, the eigenvectors of the exact Hamiltonian of the system depend on the two quantum numbers and satisfy the system of equations
\begin{equation}\label{eq:49}
  \begin{aligned}
    \hat{H}_{\mathrm{QRM}}|\psi_{n}(p)\rangle &= E_{n}(p)|\psi_{n}(p)\rangle,
    \\
    \hat{P}|\psi_{n}(p)\rangle &= p|\psi_{n}(p)\rangle,
  \end{aligned}
\end{equation}
where $p=\pm1$ defines the parity and $n=0,1,2,...$ the principle quantum numbers of the field excitations respectively.

Despite its simple form the attempt to find the exact analytical solution of the system of Eqs.~(\ref{eq:49}) with the Hamiltonian (\ref{eq:45}) is a tough challenge \cite{PhysRevLett.107.100401}. As it was demonstrated in the reference \cite{0305-4470-29-14-026}, it is possible to derive the rapidly convergent iterative expressions for the numerical computation of the eigenfunctions and eigenvalues of this system on the basis of the operator method \cite{FeranchukB2015} with any required accuracy. Moreover, even the zeroth-order approximation of the operator method provides analytical expressions for the eigenstates of the system, with relatively high accuracy in the whole range of variation of the coupling constant. We will refer to this as the uniformly available approximation. In addition, a similar approximation was derived later by other methods in the work \cite{PhysRevLett.99.173601,*PhysRevLett.99.259901}.

The following analytical expressions represent the uniformly available approximation for the eigenvalues and eigenfunctions of the quantum Rabi model (they generalize the results from \cite{0305-4470-29-14-026,PhysRevLett.99.173601,*PhysRevLett.99.259901}):
\begin{align}\label{eq:50}
  E^{(\pm)}_n &= n + \frac{1}{2} - f^2 + \frac{1}{4}\epsilon (-1)^n \left( S_{n+1,n+1}+S_{nn} \right)  \nonumber
  \\
  &\pm\frac{1}{2} \sqrt{ \left[1+\frac{\epsilon}{2}(-1)^n \left( S_{n+1,n+1}-S_{nn} \right) \right]^2 + \epsilon^2 S^2_{n+1,n}},
\end{align}
\begin{align}\label{eq:51}
  \left|\psi^{(\pm)}_n\right> &= \left\{ A^{(\pm)}_n \left|n,f \right>+B^{(\pm)}_n\left|n+1,f \right> \right\} \chi_+ \nonumber
  \\
  &+(-1)^n \hat{S} \left\{ A^{(\pm)}_n \left| n,f \right> + B^{(\pm)}_n \left| n+1,f \right> \right\} \chi_-,
\end{align}
where the coefficients $A^{(\pm)}_{n}$, $B^{(\pm)}_{n}$ and $\lambda^{(\pm)}_{n}$ read
\begin{align}\label{eq:52}
  A^{(\pm)}_n &= \frac{1}{\sqrt{2}} \frac{1}{\sqrt{1+\left(\lambda^{(\pm)}_n\right)^2}}, \quad B^{(\pm)}_n = - \lambda^{(\pm)}_n A^{(\pm)}_n, \nonumber
  \\
  \lambda^{(\pm)}_n &= \frac{n-f^2+\frac{1}{2}\epsilon (-1)^n S_{nn}-E^{(\pm)}_n}{\frac{1}{2}\epsilon (-1)^n S_{n+1,n}}.
\end{align}
For the matter of convenience in Eqs.~(\ref{eq:51}), (\ref{eq:52}) we use a different classification of the energy levels, where the two values of the quantum number $p$ are replaced with $\pm$ \cite{Feranchuk2011385}.
\begin{figure}[tb]
\includegraphics[width=0.48\linewidth]{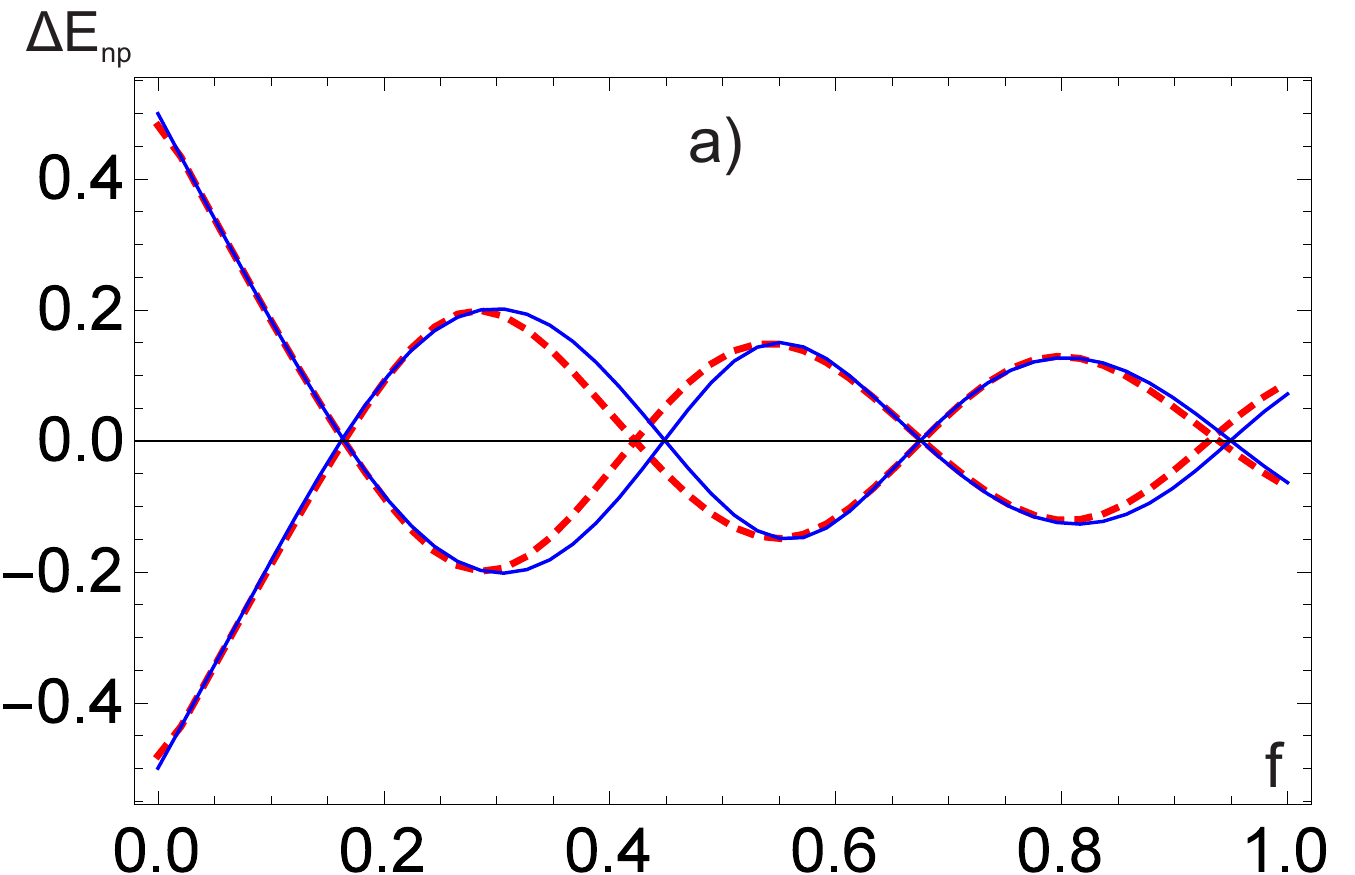}
\includegraphics[width=0.48\linewidth]{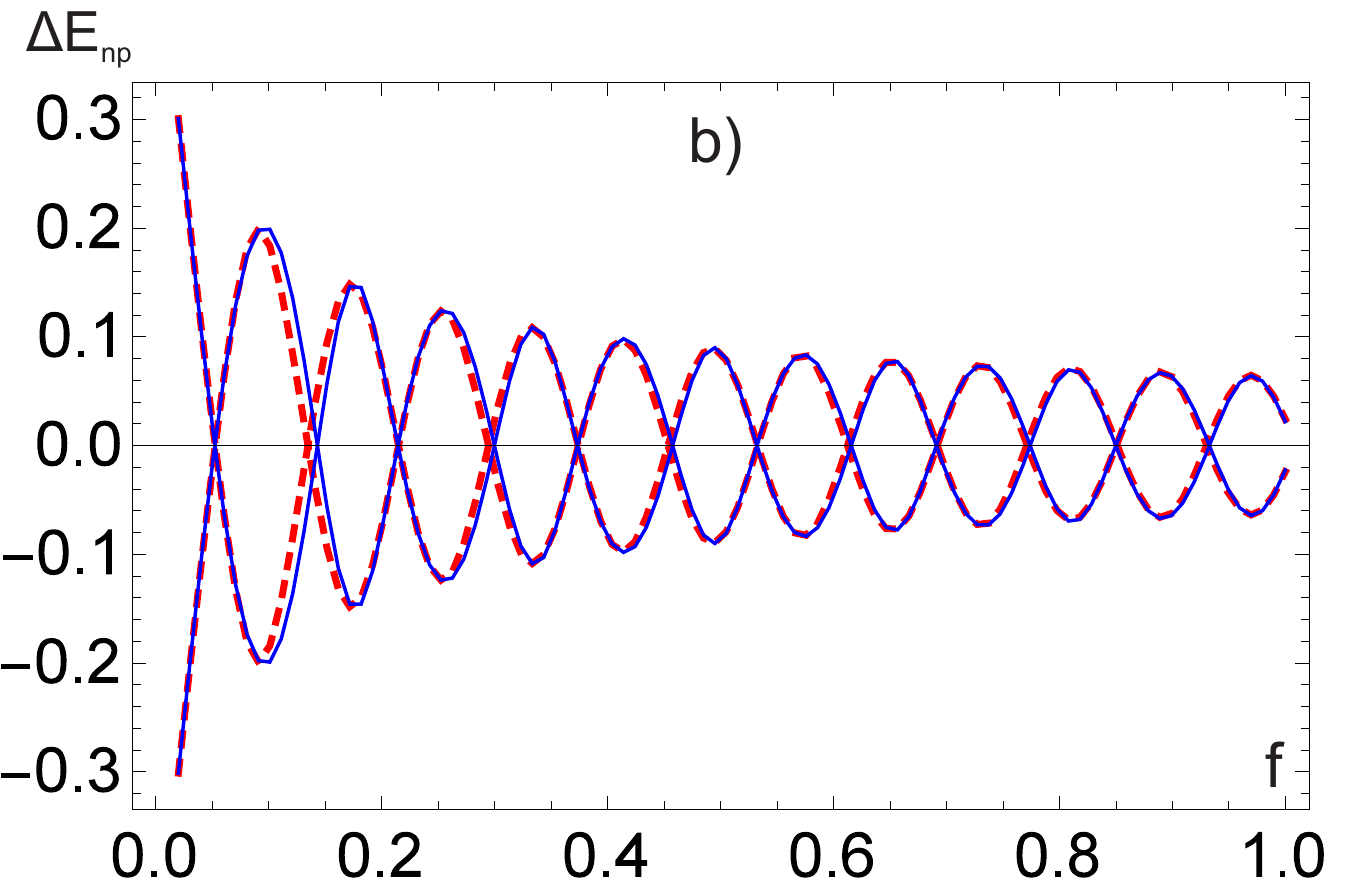}
\caption{(color online) The normalized energy levels $\Delta E_{np} = E_{np} - (n - f^2) $ of the quantum Rabi model as a function of the coupling constant $f$ in the case of exact resonance ($\omega_{0} = 1.0$) between the single-mode frequency and the atomic transition frequency:  Pane (a): $\bar n = 10$. Pane (b) $\bar n = 100$. The solid lines correspond to the uniformly available approximation and the dashed lines to the exact numerical simulations.}
\label{fig:1}
\end{figure}

Here the matrix elements of the parity operator $\hat{S}$ in the basis of the coherent states of the field $\left\{\left|n,f \right> \right\}$ ($\hat{a}|0,f\rangle = f |0,f\rangle$) are defined as \cite{FeranchukB2015}
\begin{align}\label{eq:53}
  S_{nm} (u) &= S_{mn} (u), \nonumber
  \\
  S_{nm} (u) &= (-1)^m e^{-2u^2} \sqrt{ \frac{m!}{n!} } (2u)^{n-m} L^{n-m}_m (4 u^2), \quad n \geq m,
\end{align}
where $L^{a}_{n} (x)$ are the generalized Laguerre polynomials.

The evaluation of the eigenvalues of the quantum Rabi model as a function of the coupling constant $f$ is presented in the Fig.~\ref{fig:1}, which demonstrates an excellent agreement with the exact numerical solutions of Eq.~(\ref{eq:49}) obtained as a sum of large number of terms of the series of the operator method. One can observe that all qualitative peculiarities of the exact solution are reproduced.

Let us now consider the time evolution of the quantum Rabi eigenstates. We assume that at the initial moment of time (before the interaction is switched on) the atom was in its ground state and the quantum field was in the coherent state with the coherent state parameter $\alpha \approx \sqrt{\bar n} \gg 1$, such that the initial state vector is represented as
\begin{eqnarray}\label{eq:54}
  \left| \Psi (0) \right> = |\chi_{\downarrow}\rangle \sum_{k=0}^{\infty} \frac{\alpha^k}{\sqrt{k !}} \left| k \right>e^{-\alpha^2/2}.
\end{eqnarray}
\begin{figure}[t]
  \centering
  \includegraphics[width=0.48\textwidth]{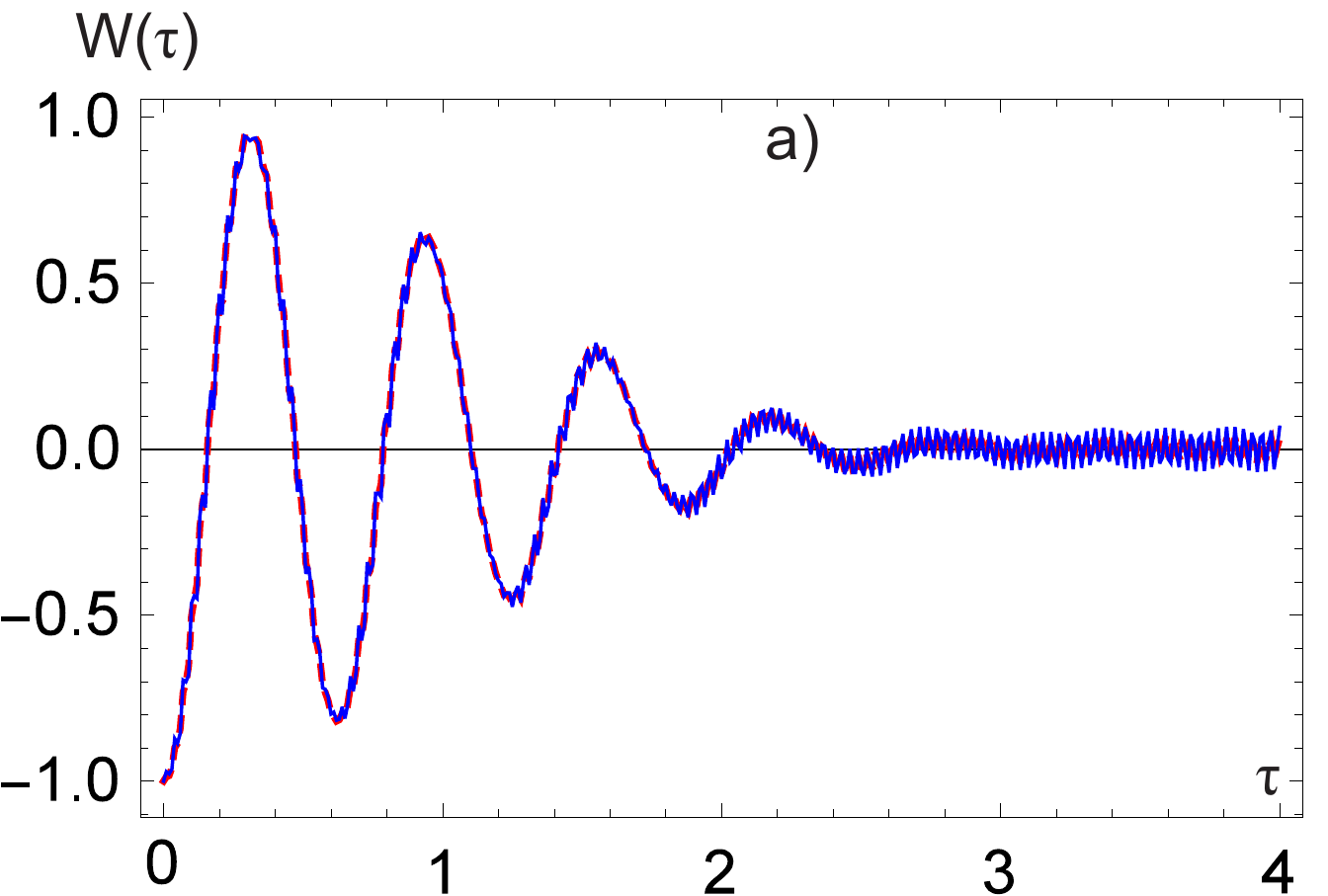}
  \includegraphics[width=0.48\textwidth]{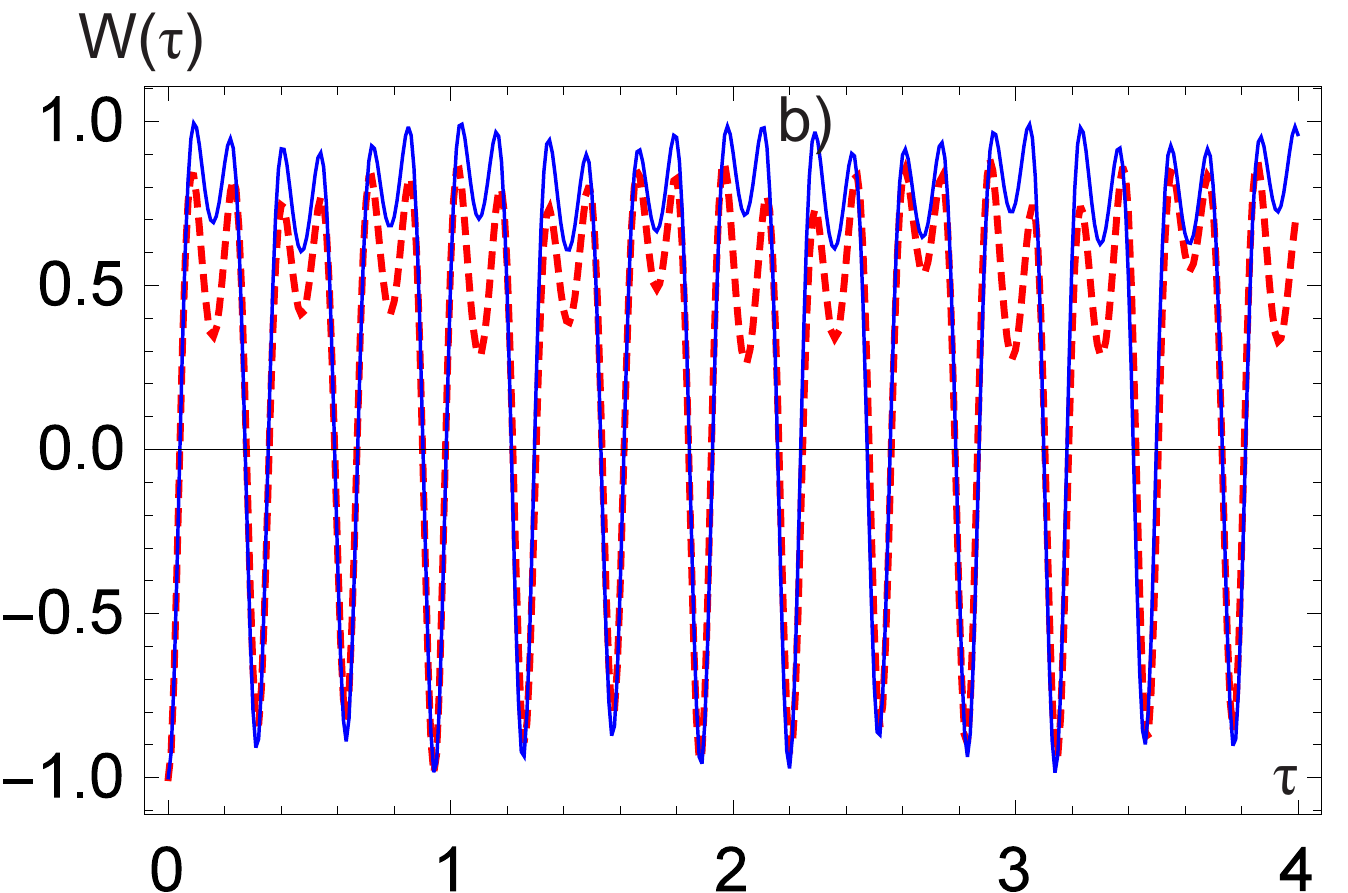}
  \\
  \includegraphics[width=0.48\textwidth]{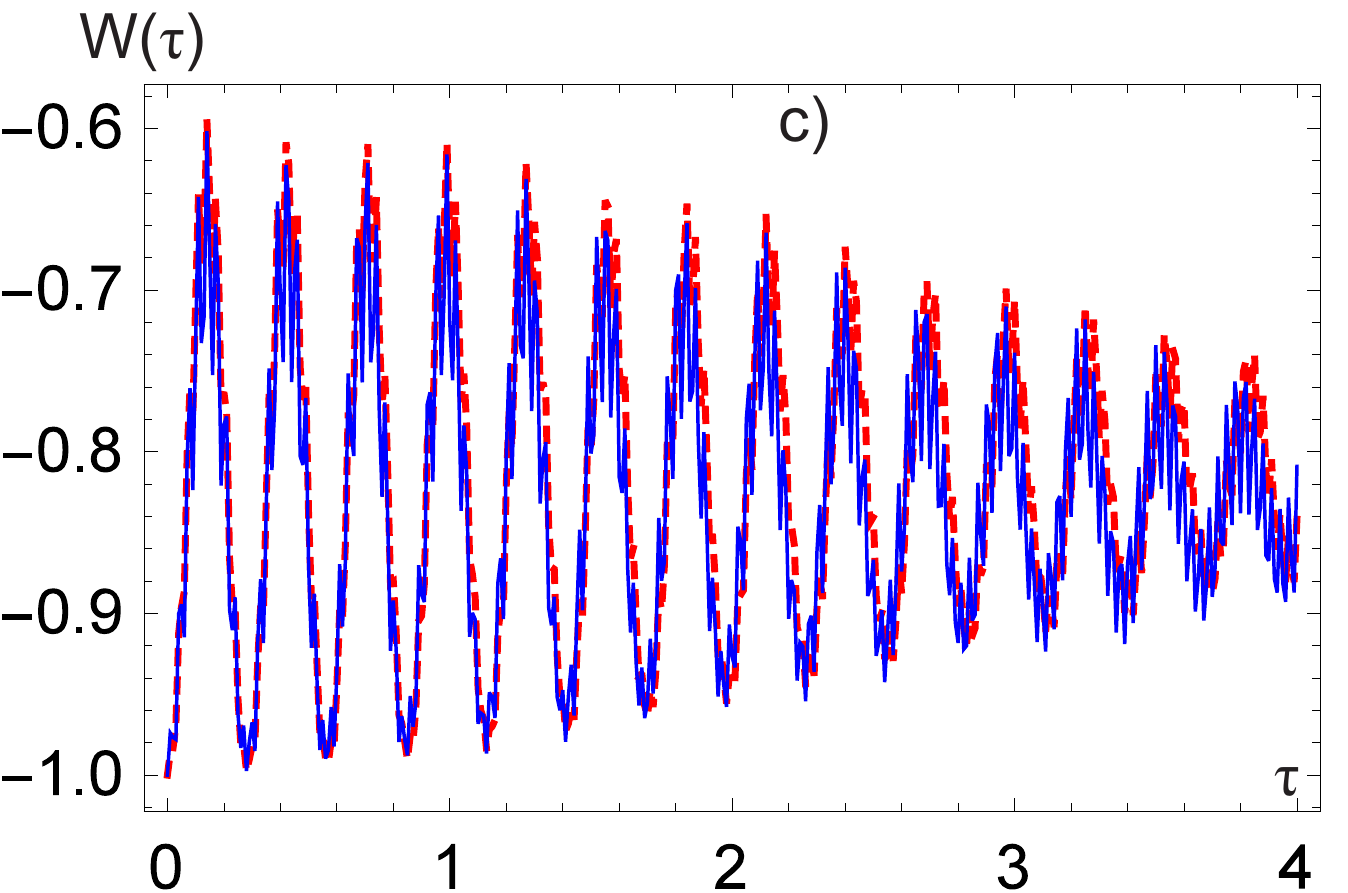}
  \includegraphics[width=0.48\textwidth]{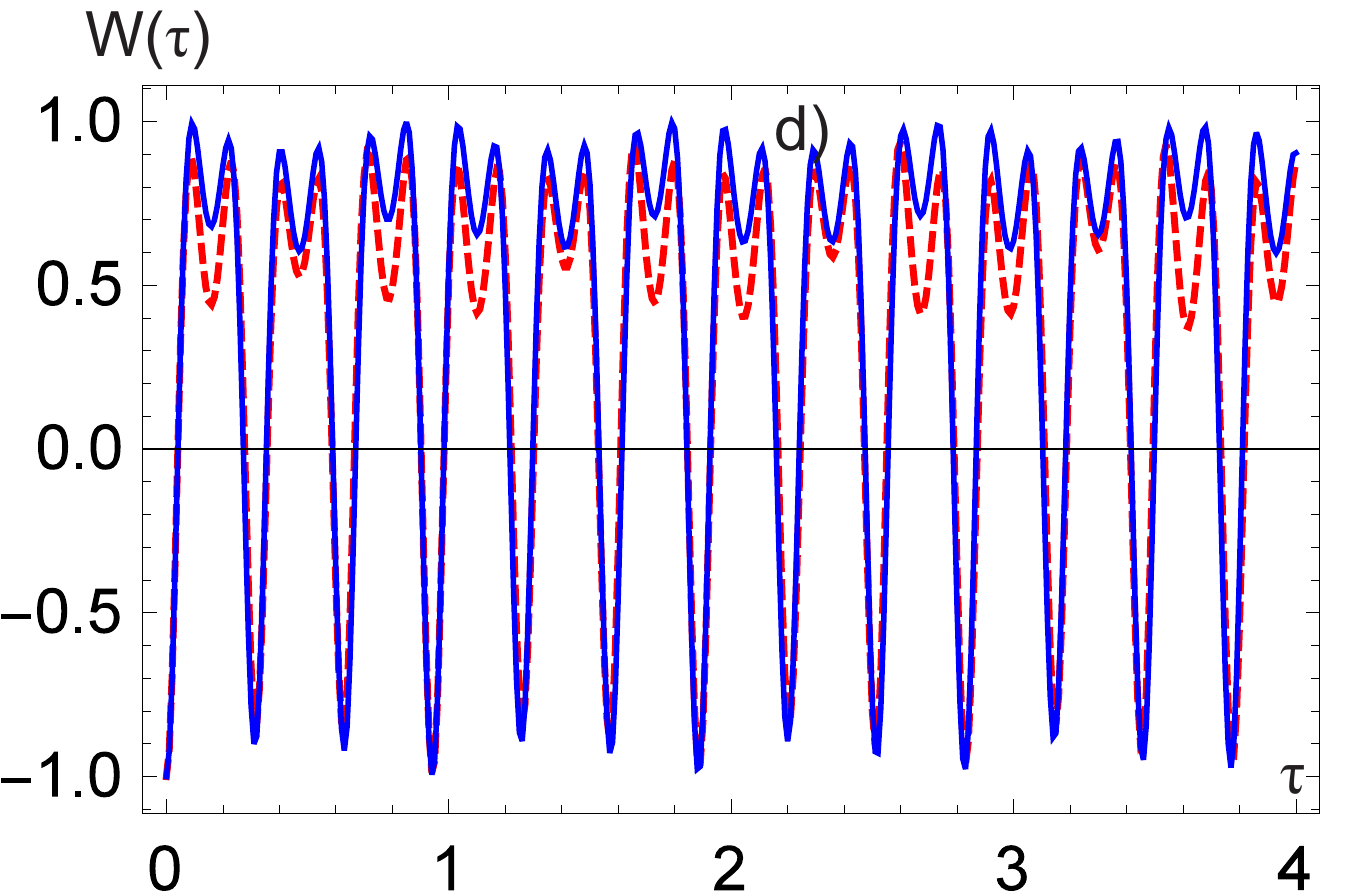}
  \caption{\label{fig:2}(color online) Population difference $W$ of a two-level system interacting with a single-mode quantized field as a function of the dimensionless time $\tau$ ($\tau = ft$). The red dashed lines represent the exact numerical evaluation, while the blue solid lines describe the uniformly available approximation. Pane~(a): The interaction constant $f=0.01$, the average photon number $\bar{n} = 25$ and the exact resonance ($\omega_{0} = 1.0$) between the single-mode frequency and the atomic transition frequency. Pane~(b): The interaction constant $f=0.1$, the average photon number $\bar n = 100$, the other parameters are the same as on Pane~(a). Pane~(c): The detuning $|\bar\epsilon - \omega_{0}| = 0.2$ between the single-mode frequency and the atomic transition frequency is introduced. The other parameters are the same as on Pane~(a). Pane~(d): The detuning $|\bar\epsilon - \omega_{0}| = 0.2$ between the single-mode frequency and the atomic transition frequency is introduced. The other parameters are the same as on Pane~(b).}
\end{figure}

The time evolution of the coherent state (\ref{eq:54}) is defined by the following expansion:
\begin{eqnarray}
\label{eq:55}
    \left| \Psi (t) \right> = \sum^\infty_{n=0} \sum_{r=\pm} C_{nr} \left|\psi^{(r)}_n\right>
        e^{-i E^{(r)}_n t},
\end{eqnarray}
where the coefficients $C_{nr}$ of the expansion are calculated from the initial condition Eq.~(\ref{eq:54}).

For the following we need the density matrix
\begin{eqnarray}
\label{eq:56}
    \hat{\rho}_{\mathrm{a}} (t) = \Tr_{f} \left\{ \left| \Psi (t) \right> \left< \Psi (t) \right| \right\} =
        \sum^{\infty}_{k=0} \left< k,f \right| \hat{\rho}_{\mathrm{a}} \left| k,f \right> = \left( \begin{array}{cc}
                                                                                        \rho_{\uparrow\uparrow} & \rho_{\uparrow\downarrow}
                                                                                        \\
                                                                                        \rho^*_{\uparrow\downarrow} & \rho_{\downarrow\downarrow}
                                                                                      \end{array} \right),
\end{eqnarray}
such that the population difference is calculated as \cite{ScullyB1997}
\begin{eqnarray}
\label{eq:57}
    W(t) = \rho_{\uparrow\uparrow} - \rho_{\downarrow\downarrow}.
\end{eqnarray}

The exact expressions (\ref{eq:55})--(\ref{eq:57}) based on the stationary states (\ref{eq:49}) are calculated within the framework of the operator method. The number of intermediate states $\Delta n$ that is needed to perform the summation in Eq.~(\ref{eq:55}) and the dimension of the matrices $\Delta m$ for the numerical solution of (\ref{eq:49}) are defined by the dimensionless field amplitude $\Delta n \sim \Delta m \sim \alpha$. However, we present below the analytical approximation for the evolution operator on the basis of Eqs.~(\ref{eq:50})--(\ref{eq:51}).

For this purpose we substitute Eqs.~(\ref{eq:50}) and (\ref{eq:51}) into (\ref{eq:55})--(\ref{eq:57}) and obtain for the population difference
\begin{align}\label{eq:58}
  W(t) = 2 \re\left\{ \sum^{\infty}_{k=0} \sum^{\infty}_{n=0} \sum_{r=\pm} \sum_{q=\pm} D^{rq}_{kn}e^{-i \left( E^{(r)}_n - E^{(q)}_k \right) t} \right\},
\end{align}
where
\begin{align}\label{eq:59}
  D^{rq}_{kn} &\equiv (-1)^n C_{nr} C^*_{kq} \nonumber
  \\
  &\times\left[ A^{(q)}_{k} \left( A^{(r)}_n S_{kn} + B^{(r)}_n S_{k,n+1} \right) + B^{(q)}_k \left( A^{(r)}_n S_{k+1,n} + B^{(r)}_n S_{k+1,n+1} \right) \right],
\end{align}
with the evolution coefficients
\begin{align}\label{eq:60}
  C_{nr} &= \frac{1}{\sqrt{2}} e^{-\frac{1}{2} \left( \alpha + f \right)^2} \left\{ \frac{\left( \alpha + f \right)^n}{\sqrt{n!}} \left(A^{(r)}_n + B^{(r)}_n \frac{\left( \alpha + f \right)}{\sqrt{n+1}} \right)-
  (-1)^n \gamma_{nr} \right\}, \nonumber
  \\
  \gamma_{nr} &\equiv \sum^{\infty}_{m=0} \frac{ \left(\alpha+f \right)^m}{\sqrt{m!}} \left(A^{(r)}_n S_{mn}
        + B^{(r)}_n S_{m,n+1} \right).
\end{align}

The comparison of the time evolution of the population difference $W$ obtained via Eq.~(\ref{eq:58}) (the uniformly available approximation) and the summation of a large number terms of the operator method are presented in Fig.~\ref{fig:2}. The Fig.~\ref{fig:2} demonstrates an excellent agreement in the whole range of $f\sqrt{\bar{n}}$, as in the case of the exact resonance ($\omega_{0} = 1.0$) between the single-mode frequency and the atomic transition frequency, as well as when the detuning is introduced.

\begin{figure}[t]
  \centering
  \includegraphics[width=0.48\textwidth]{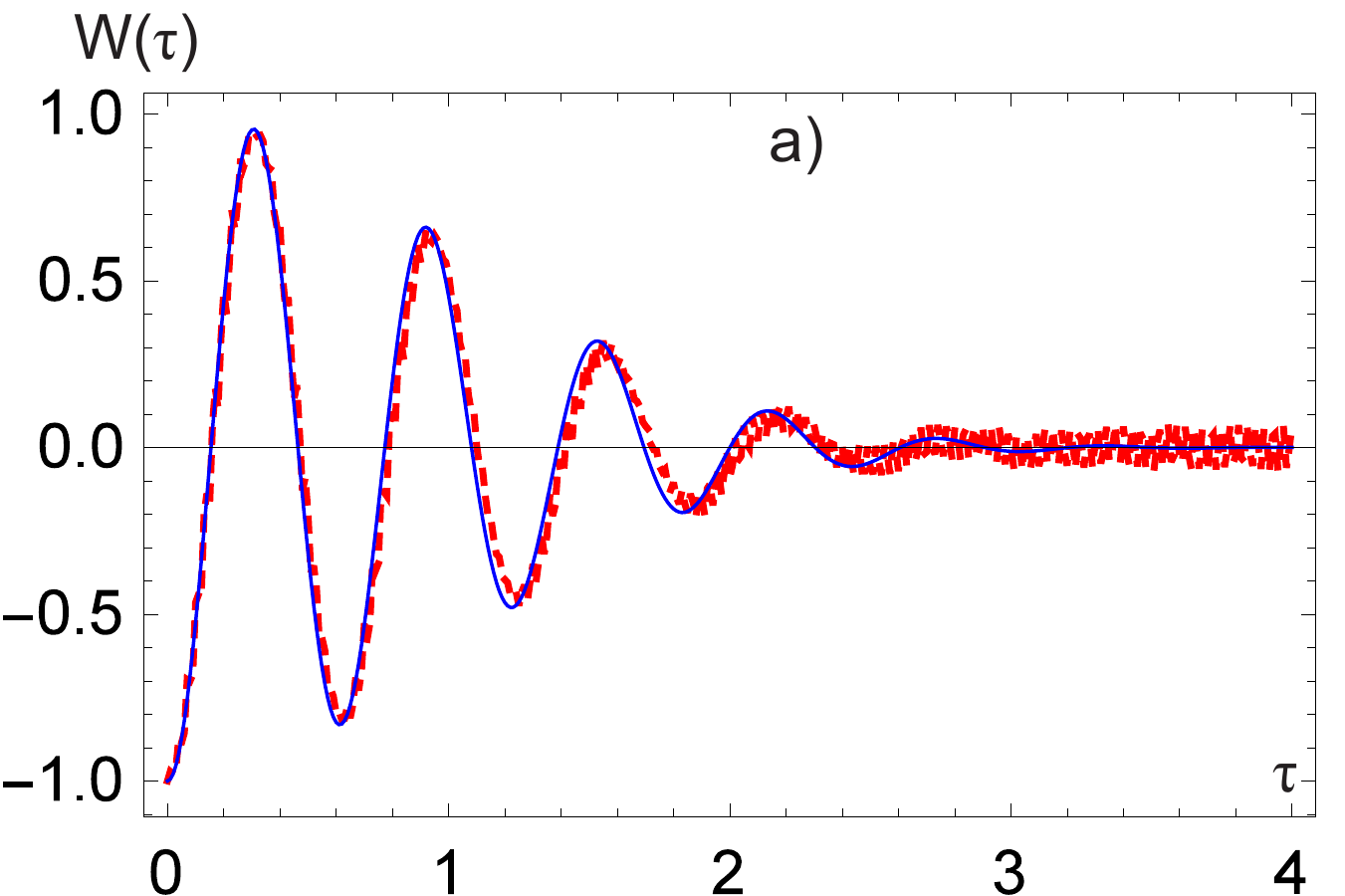}
  \includegraphics[width=0.48\textwidth]{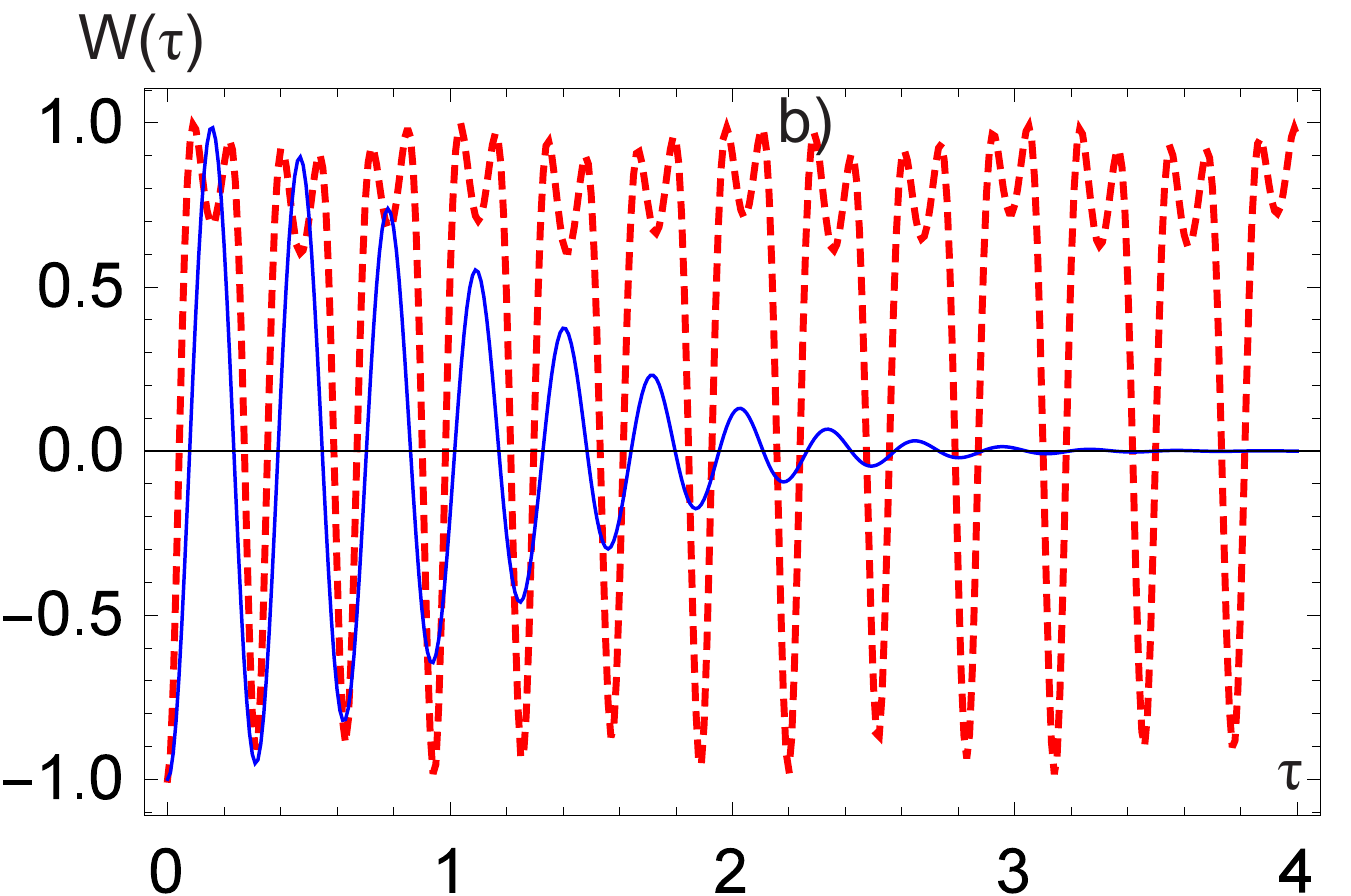}
  \\
  \includegraphics[width=0.48\textwidth]{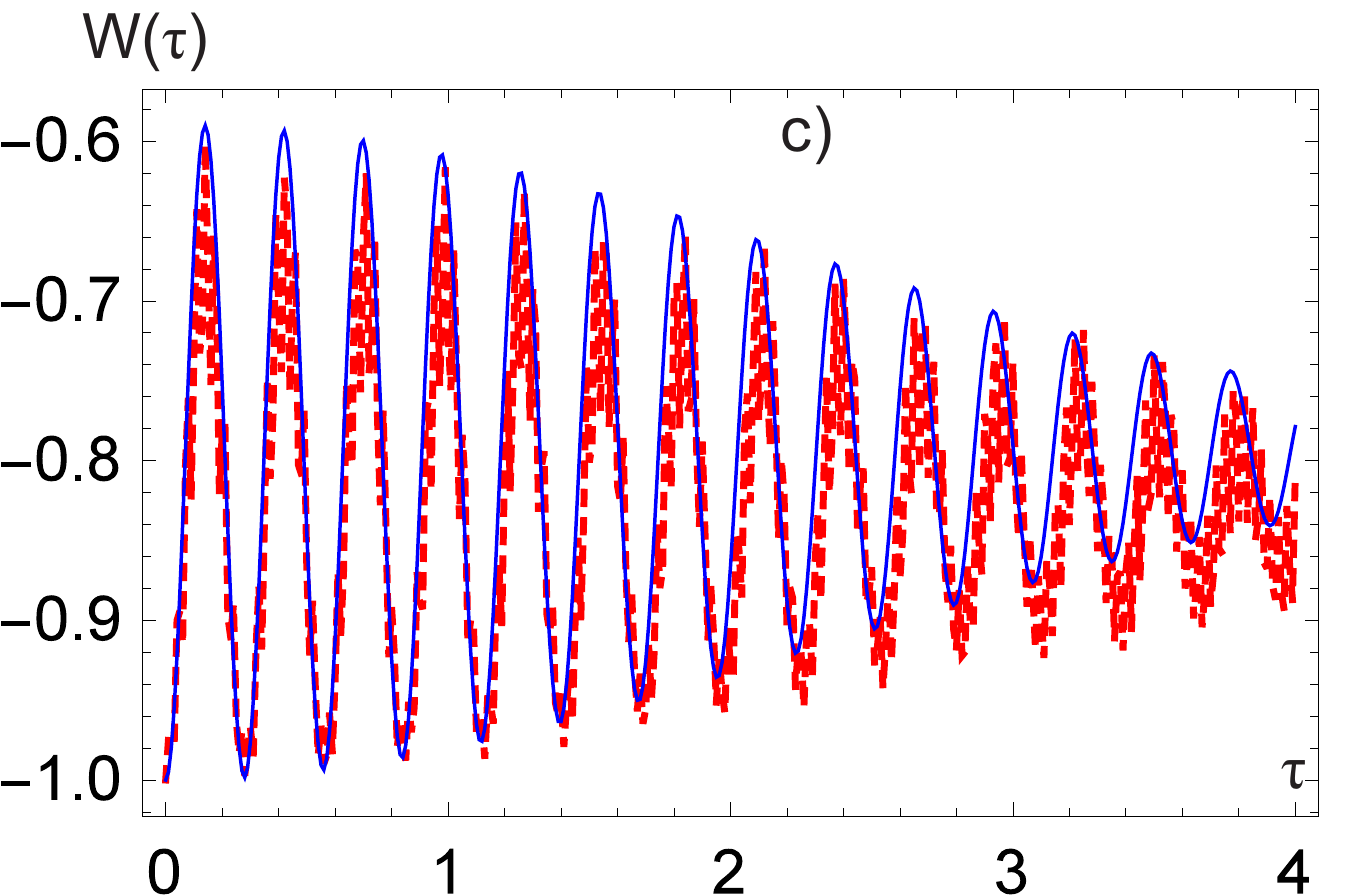}
  \includegraphics[width=0.48\textwidth]{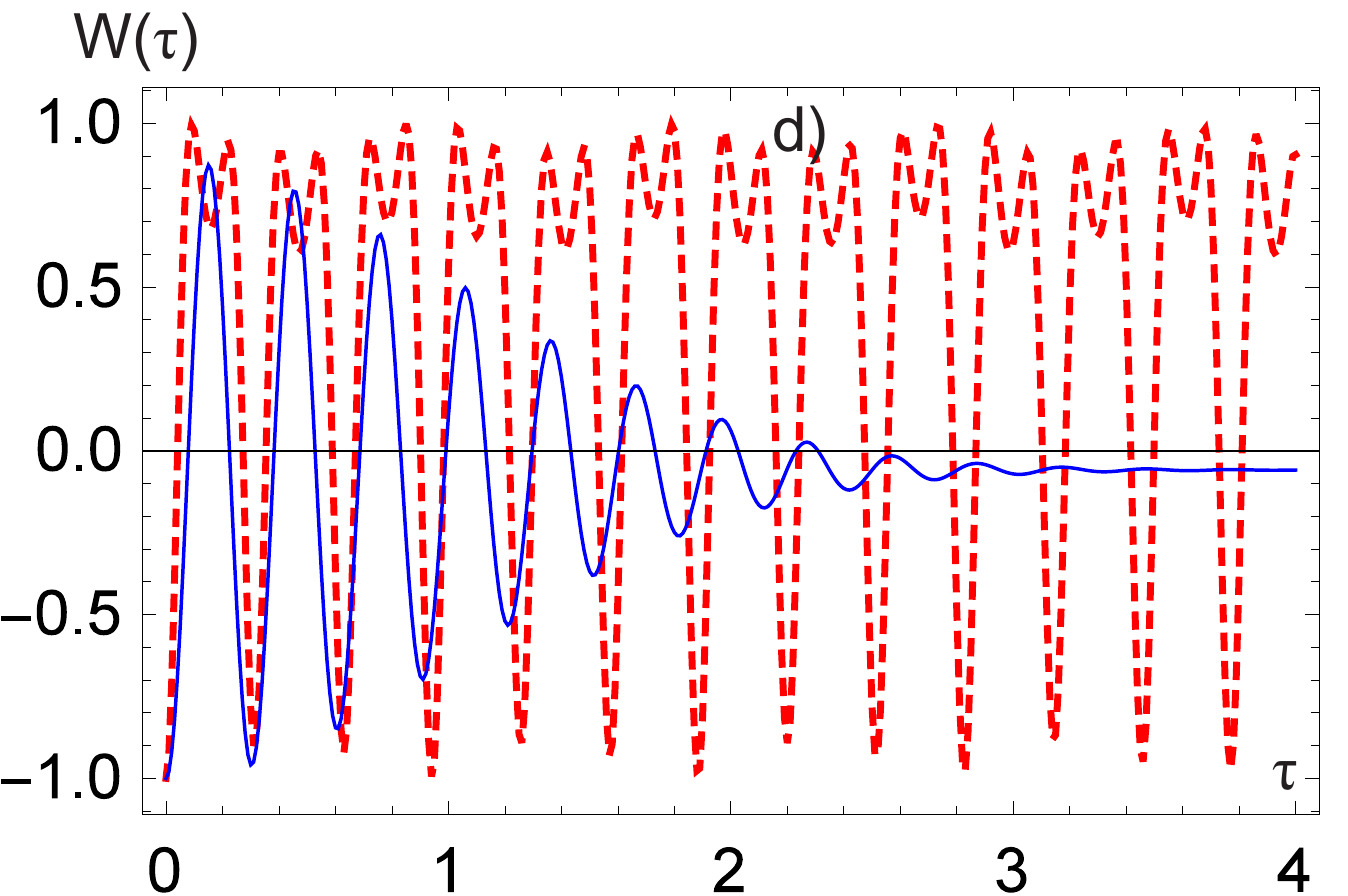}
  \caption{\label{fig:3} (color online) Population difference $W$ of a two-level system interacting with a single-mode quantized field as a function of the dimensionless time $\tau$ ($\tau = ft$). The red dashed lines represent the uniformly available approximation, while the blue solid lines describe the rotating wave approximation. Pane~(a): The interaction constant $f=0.01$, the average photon number $\bar{n} = 25$ and the exact resonance ($\omega_{0} = 1.0$) between the single-mode frequency and the atomic transition frequency. Pane~(b): The interaction constant $f=0.1$, the average photon number $\bar n = 100$, the other parameters are the same as on Pane~(a). Pane~(c): The detuning $|\bar\epsilon - \omega_{0}| = 0.2$ between the single-mode frequency and the atomic transition frequency is introduced. The other parameters are the same as on Pane~(a). Pane~(d): The detuning $|\bar\epsilon - \omega_{0}| = 0.2$ between the single-mode frequency and the atomic transition frequency is introduced. The other parameters are the same as on Pane~(b).}
\end{figure}

The comparison of the evolution of the population difference $W$ as a function of time within and beyond the rotating wave approximation is presented in Fig.~\ref{fig:3}. As can be seen from this figure, the evolution of the system in the rotating wave approximation is substantially different from the one beyond the rotating wave approximation in the regime of the strong field. Moreover, even the qualitative peculiarities can not be reproduced.

Nevertheless, the actual usability of the expressions defined via Eqs.~(\ref{eq:58})--(\ref{eq:60}) for the calculation of the observable quantities of the system is still complicated. However, in the practically important case of the strong field ($\bar n \gg 1$ and $f\sqrt{\bar n} \gg 1$), it is possible to carry out an analytical summation over the intermediate states in Eq.~(\ref{eq:58}) and obtain a compact expression for the population difference as a function of time. The further derivations are based on the implementation of the following relation for the sum over matrix elements:
\begin{align}\label{eq:61}
  \sum^{\infty}_{m=0} S_{mn} \approx (-1)^n + O \left[ n^{-1} \right],
\end{align}
which can be derived from the asymptotic behavior of the Laguerre polynomials.

With the help of Eq.~(\ref{eq:61}) one can simplify the corresponding expression for $\gamma_{nr}$ in Eq.~(\ref{eq:60})
\begin{align}\label{eq:62}
  \gamma_{nr} \approx (-1)^n \left\{ A^{(r)}_n \frac{\left(\alpha + f \right) ^n}{\sqrt{n!}} - B^{(r)}_n \frac{\left(\alpha + f \right) ^{n+1}}{\sqrt{(n+1)!}} \right\},
\end{align}
and obtain the following representation of the evolution coefficients
\begin{align}\label{eq:63}
  C_{nr} \approx \sqrt{2} e^{-\frac{1}{2} \left( \alpha + f \right)^2} \frac{\left(\alpha + f \right) ^{n+1}}{\sqrt{(n+1)!}} B^{(r)}_n.
\end{align}

The coefficients $A^{(\pm)}_n$ and $B^{(\pm)}_n$ are smooth functions of the parameter $n$. Moreover, the main contribution to the sum in Eq.~(\ref{eq:58}) is given by the small region near the value $\bar n=\bar n_0=\alpha ^2$ with variance $\mathcal{D}n \sim \alpha$. Consequently, these coefficients can be evaluated in the central point $n = n_{0}$ and removed from the sum. As a result, Eq.~(\ref{eq:58}) is transformed into the form
\begin{align}\label{eq:64}
  W(t) \approx 4 e^{-\alpha ^2} \sum_{r=\pm} \sum_{q=\pm} \sum^{\infty}_{n=0} \sum^{\infty}_{k=0} \xi_n \tilde{D}^{rq}_{kn} \cos \Omega^{rq}_{kn} t,
\end{align}
where we introduced the abbreviations
\begin{equation}\label{eq:65}
  \begin{aligned}
      \Omega^{rq}_{kn} &\equiv E^{(r)}_k - E^{(q)}_n; \quad \xi_n \equiv (-1)^n \frac{ \left( \alpha + f \right)^{2(n+1)}}{(n+1)!};
  \\
  \tilde{D}^{rq}_{kn} &\equiv A^{(q)}_{n_0}A^{(r)}_{n_0}B^{(q)}_{n_0}B^{(r)}_{n_0}S_{nk} + A^{(q)}_{n_0}B^{(r)}_{n_0}B^{(r)}_{n_0}B^{(q)}_{n_0+1}S_{n+1,k}
  \\
  &+A^{(r)}_{n_0}B^{(q)}_{n_0}B^{(r)}_{n_0}B^{(q)}_{n_0}S_{n,k+1} + B^{(r)}_{n_0}B^{(r)}_{n_0}A^{(q)}_{n_0+1}B^{(q)}_{n_0+1}S_{n+1,k+1}.
  \end{aligned}
\end{equation}

The further simplification can be achieved with the use of an asymptotic representation of the Laguerre polynomials in the limit of $n \gg 1$, which is expressed through the Bessel functions \cite{AbramowitzB1964}
\begin{align}\label{eq:66}
  L^{(a)}_n (x) \approx \left( \frac{n}{x} \right) ^ \frac{a}{2} J_a (2 \sqrt{nx}).
\end{align}
Consequently, the matrix elements of the operator $\hat S$ take the form 
\begin{align}\label{eq:67}
  S'_{nk} (f) = (-1)^n e^{-2f^2} e^{- \frac{k^2}{4n}} J_k (4f\sqrt{n})
\end{align}
where the prime implies the fact that the matrix element $S_{mn}$ in Eq.~(\ref{eq:53}) coincides with the $S'_{n,m-n}$, i.e., the shift of the index $k = m - n$ is performed. In the following we will use only the asymptotic elements and consequently the prime will be omitted below.

The double sum over $r$ and $q$ in Eq.~(\ref{eq:62}) can be separated into groups with $r=q$ and $r=-q$ respectively. The corresponding asymptotic expansion of the frequencies in Eq.~(\ref{eq:65}) yields
\begin{align}\label{eq:68}
  \Omega^{r,r}_{kn} &\approx (n-k); \quad \Omega^{r,-r}_{kn} \approx (n-k) +r\beta_n; \nonumber
  \\
  \beta_n &\equiv \sqrt{ \left[ 1- \epsilon e^{-2f^2} J_0 (4f \sqrt{n}) \right] ^2 +\epsilon^2 e^{-4f^2} J^2_1(4f\sqrt{n})}.
\end{align}

At last, with the help of Eqs.~(\ref{eq:66}), (\ref{eq:67}), (\ref{eq:68}) and relations for the Bessel functions \cite{AbramowitzB1964}
\begin{align}\label{eq:69}
  J_0 (z) + 2 \sum^{\infty}_{k=1} J_{2k} (z) \cos[2k \theta] = \cos[z \sin \theta]; \quad \sum^{\infty}_{-\infty} J_k (z) t^k = e^{\frac{1}{2} \left(t - \frac{1}{t} \right) z}
\end{align}
the summation over the index $k$ can be performed.

Finally, in order to reach the answer the summation over $n$ is replaced by an integration. Then with the help of the saddle point method \cite{MorseB1953}, one can derive the approximate analytical equation for the population difference of the two-level system interacting with a single-mode quantized field, which is valid in the limit of $\alpha \gg 1$ and $f \alpha \gg 1$
\begin{eqnarray}\label{eq:70}
  W(t) \approx -\cos \left(4f \alpha \sin t \right)
\end{eqnarray}
\begin{figure}[tb]
  \includegraphics[width=0.48\linewidth]{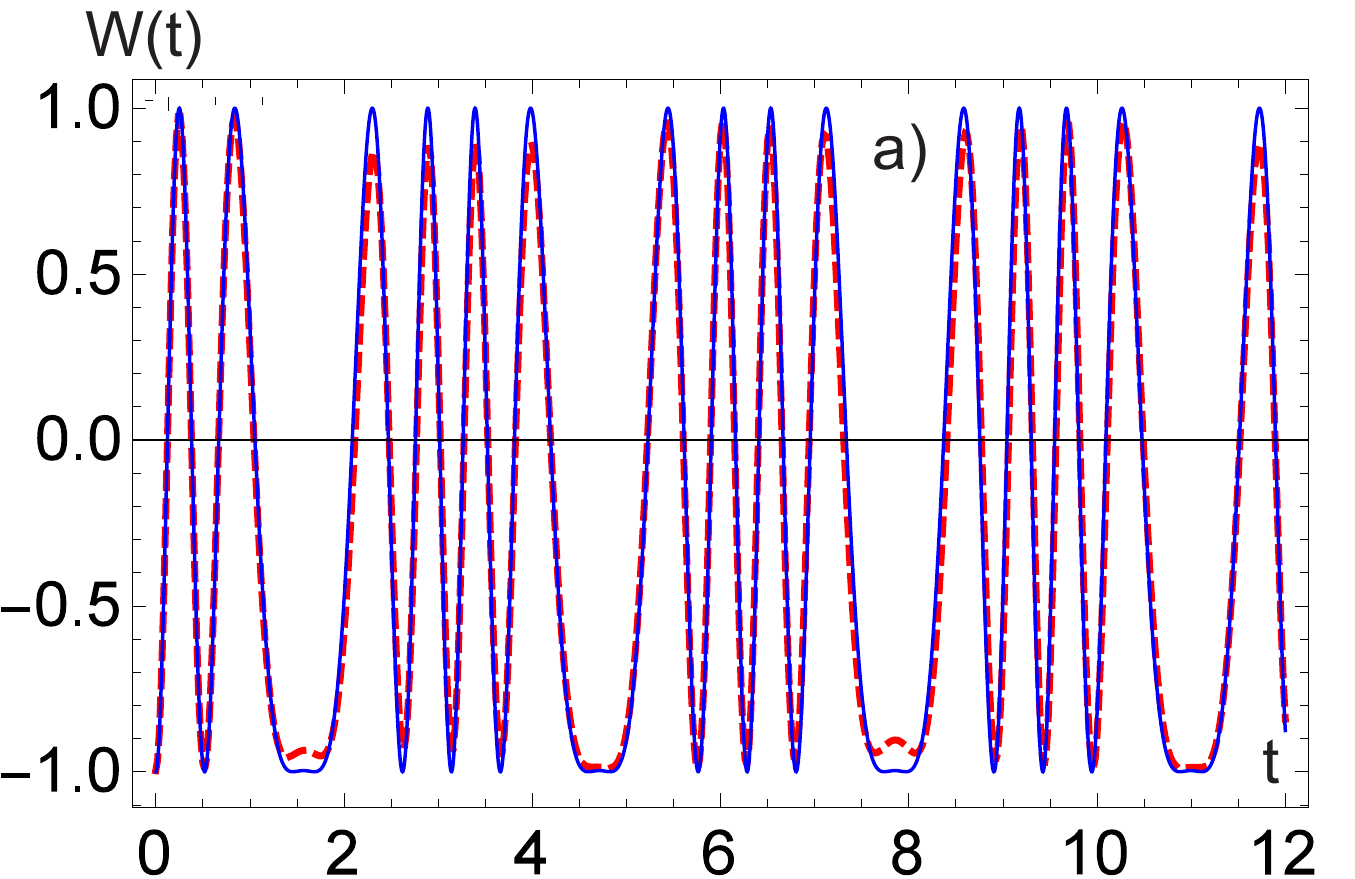}
  \includegraphics[width=0.48\linewidth]{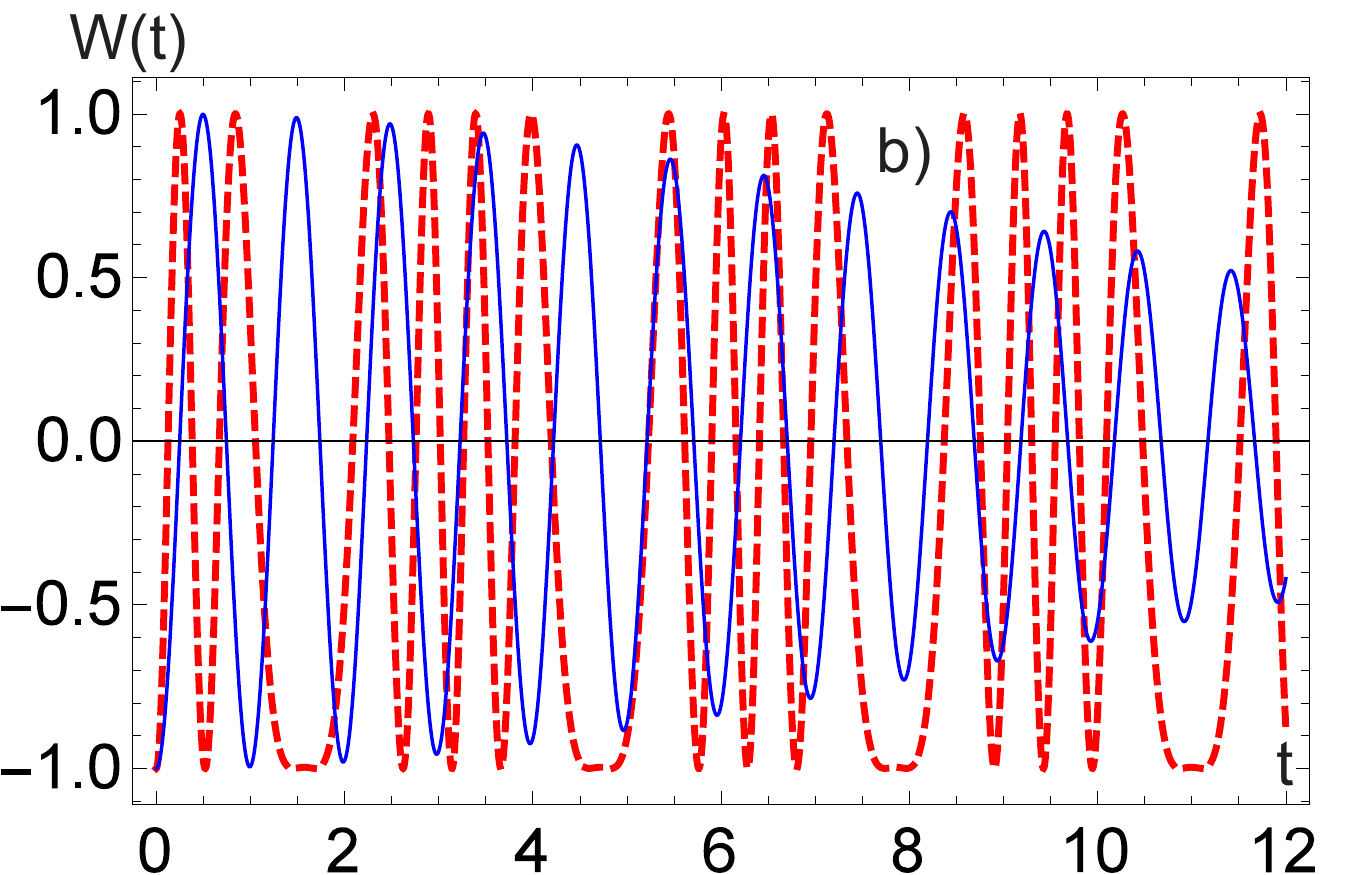}
  \caption{(color online) Population difference $W$ of a two-level system interacting with a single-mode quantized field as a function of time (arb. u.) in the case of exact resonance ($\omega_{0} = 1.0$) between the single-mode frequency and the atomic transition frequency, $\bar n = 10^3$ and $f = 0.1$. Pane (a): The dashed red line represents the exact numerical solution (\ref{eq:58}) and the blue solid line the approximate relation (\ref{eq:70}). Pane (b): The red dashed line represents the approximate relation (\ref{eq:70}) and the blue solid line the inverse population in the rotating wave approximation \cite{ScullyB1997}.}
  \label{fig:4}
\end{figure}

The comparison of the simple analytical expression Eq.~(\ref{eq:70}) with the exact numerical simulation of the full evolution operator in Fig.~\ref{fig:4} demonstrates an excellent agreement for the population difference of the two-level system interacting with a single-mode quantized field. At the same time, the use of the rotating wave approximation does not even reproduce the qualitative peculiarities of the system evolution.

\section{Conclusion}
\label{sec:conclusion}
In this paper we have investigated the applicability of the two major approximations which are most commonly employed in the study of the quantum Rabi model, namely the description of a resonant cavity mode as a single-mode quantized field and the use of the rotating wave approximation. We have demonstrated that in a real cavity with finite Q-factor a finite distribution of modes has to be considered. Consequently, due to energy dissipation processes a two-level system interacts with a field wave packet, which is centered at the resonant cavity eigenmode. Starting from the Schr\"{o}dinger equation describing the interaction between the two-level system and the multi-mode quantized field we performed the canonical transformation of the field variables, and consequently divided this interaction Hamiltonian into two parts. The first part describes the interaction between the atom and the single-collective field mode, while the second describes the interaction with the fluctuations. Afterwards, we have shown that the interplay between the energy of the fluctuations and the interaction energy between the atom and the collective-field mode defines the applicability conditions of the single-mode approximation. We have found that in the case when the energy density $w$ in the resonant cavity mode is larger than the critical energy density
\begin{equation*}
   w > w_{\mathrm{c}} = \frac{m_{\mathrm{e}}\omega_{0}^{3}}{49 e_{0}^{2}} \approx \frac{5.7\cdot10^{10}}{\lambda_{0}^{3}[\mathrm{nm}]}\left[\frac{\mathrm{J}}{\mathrm{cm}^{3}}\right],
\end{equation*}
the field can be described with good accuracy as a single-mode.

After establishing this condition we switched to the analysis of the stationary states and the time evolution of the system beyond the rotating wave approximation. It was shown that in this case the integrals of motion are different from the ones of the rotating wave approximation, namely the operator of the combined parity $\hat P = \hat \sigma_{3}\exp(i\pi\hat A^{\dag}\hat A)$ commutes with the Hamiltonian of the quantum Rabi model. Thereafter, we were able to determine simple analytical expressions that allow one to calculate the spectrum of the system with arbitrary required accuracy. Moreover, our results are valid in the whole range of variation of the coupling constant. This was proven by comparing the exact numerical simulations with the approximate analytical formulas. Furthermore, we analyzed the time evolution of the system numerically, assuming that the field at the initial moment of time was in the coherent state. In the experimentally important regime of large photon occupation numbers we derived an  extremely simple analytical formula for the description of the time dependence of the inverse population
\begin{equation*}
  W(t) \approx -\cos \left(4f \alpha \sin t \right),
\end{equation*}
which is in excellent agreement with the exact numerical simulations. Finally, it was shown that in the regime of $n \gg 1$ the use of the rotating wave approximation does not even reproduce the qualitative peculiarities of the system evolution, i.e., the suppression of collapses in the collapse--revival effect and the qualitative changes of the time evolution of the population difference in comparison with the Rabi oscillations.

\section{Acknowledgment}
The authors are grateful to C. H. Keitel and S. M. Cavaletto for the useful discussions. AVL and IDF would like to thank Alexander von Humboldt Foundation, Research Group Linkage Program for the financial support.

\bibliography{rabi}

\end{document}